\documentclass{article} 
\usepackage{iclr2026_conference,times}

\usepackage{amsmath,amsfonts,bm}

\def\eqref#1{equation~\ref{#1}}

\def\1{\bm{1}}

\DeclareMathAlphabet{\mathsfit}{\encodingdefault}{\sfdefault}{m}{sl}
\SetMathAlphabet{\mathsfit}{bold}{\encodingdefault}{\sfdefault}{bx}{n}

\usepackage{fancyvrb}
\usepackage{fvextra}
\usepackage{hyperref}
\usepackage{url}
\usepackage{svg}  
\usepackage{siunitx}
\svgsetup{inkscapelatex=true, inkscapearea=page}
\usepackage{graphicx} 
\usepackage{tabularx}
\usepackage{pifont}
\usepackage{longtable}
\usepackage{adjustbox} 
\usepackage{booktabs}
\usepackage{caption}
\usepackage{float}
\usepackage[utf8]{inputenc} 
\usepackage[T1]{fontenc}    
\usepackage{hyperref}       
\usepackage{url}            
\usepackage{booktabs}       
\usepackage{amsfonts}       
\usepackage{nicefrac}       
\usepackage{microtype}      
\usepackage{xcolor}         %
\usepackage[table]{xcolor}
\usepackage{booktabs,tabularx,seqsplit}

\usepackage{bm}
\usepackage{multirow}
\usepackage{graphicx}
\usepackage{algorithmicx}
\usepackage{algorithm}
\usepackage{algorithm,algorithmicx,algpseudocode}
\usepackage{algpseudocode}
\usepackage{wrapfig}
\usepackage{amsmath}
\usepackage{amssymb}
\usepackage{booktabs}
\usepackage{array} 
\usepackage{adjustbox}
\usepackage{bbding}
\usepackage{ulem}
\usepackage{makecell}
\usepackage{enumitem}
\usepackage{colortbl}
\usepackage{wrapfig}
\usepackage{titletoc}
\usepackage{minitoc}
\usepackage{changepage}

\usepackage[capitalize,noabbrev]{cleveref}
\usepackage{xcolor}
\usepackage{colortbl}
\usepackage{multirow}
\usepackage{bm}
\usepackage{tcolorbox}

\usepackage{caption}
\usepackage{subcaption}

\usepackage{fvextra}

\fvset{
   fontsize=\small,
   breaklines=true,          
   breakanywhere=true,        
   numbersep=4pt,
}

\definecolor{lightgray}{gray}{0.9}
\definecolor{lightblue}{rgb}{0.8,0.9,1}
\definecolor{lightcoral}{rgb}{0.98, 0.9, 0.9}

\definecolor{codegreen}{rgb}{0,0.3,0.6}
\definecolor{codegray}{rgb}{0.5,0.5,0.5}
\definecolor{codepurple}{rgb}{0.58,0,0.82}
\definecolor{backcolour}{rgb}{0.95,0.95,0.92}
\definecolor{darkblue}{rgb}{0.0,0.0,0.66}   
\hypersetup{colorlinks=true,linkcolor=darkblue,citecolor=darkblue}

\title{VCWorld: A Biological World Model for Virtual Cell Simulation}

\author{
  Zhijian Wei\textsuperscript{1,2,}\thanks{These authors contributed equally to this work. $^\dagger$ Corresponding author.} \quad
  Runze Ma\textsuperscript{1,2,}\footnotemark[1] \quad
  Zichen Wang\textsuperscript{1} \quad
  Zhongmin Li\textsuperscript{1} \\
  \vspace{0.1cm}
  \textbf{Shuotong Song\textsuperscript{1}} \quad
  \textbf{Shuangjia Zheng\textsuperscript{1, $\dagger$}} \\
  \textsuperscript{1} Shanghai Jiao Tong University \textsuperscript{2} NeoLife AI \\
  \texttt{\{realman\_01, runze.ma, zichen\_w, shuangjia.zheng\}@sjtu.edu.cn}
}

\date{}

\iclrfinalcopy

\begin{document}

\maketitle

\begin{abstract}

Virtual cell modeling aims to predict cellular responses to perturbations. Existing virtual cell models rely heavily on large-scale single-cell datasets, learning explicit mappings between gene expression and perturbations. Although recent models attempt to incorporate multi-source biological information, their generalization remains constrained by data quality, coverage, and batch effects. More critically, these models often function as black boxes, offering predictions without interpretability or consistency with biological principles, which undermines their credibility in scientific research. To address these challenges, we present VCWorld, a cell-level white-box simulator that integrates structured biological knowledge with the iterative reasoning capabilities of large language models to instantiate a biological world model. VCWorld operates in a data-efficient manner to reproduce perturbation-induced signaling cascades and generates interpretable, stepwise predictions alongside explicit mechanistic hypotheses. In drug perturbation benchmarks, VCWorld achieves state-of-the-art predictive performance, and the inferred mechanistic pathways are consistent with publicly available biological evidence. Our code is publicly available at \url{https://github.com/GENTEL-lab/VCWorld}.
\end{abstract}

\section{Introduction}

Cells, the fundamental units of life, maintain organismal function and homeostasis through a complex interplay of biochemical processes~\citep{alberts2002cell}. A central challenge in modern biology and drug discovery is to understand and predict how cells respond to external perturbations, such as drug treatments or genetic edits~\citep{liberali2015single,lotfollahi2019scgen}. The ability to forecast these cellular state changes \textit{in silico} would not only illuminate the mechanisms of complex diseases but also accelerate the development of novel therapeutics by reducing the time and cost of experimental screening~\citep{del2010diseases}. The concept of the virtual cell, which leverages computational models to simulate cellular behavior, has emerged as a promising paradigm to address this challenge~\citep{bunne2024build}.

Recent advances in deep learning, coupled with significant progress in single-cell sequencing technologies, have spurred the development of various virtual cell models~\citep{lopez2018deep,lotfollahi2023predicting,adduri2025predicting, tang2025cellforge,klein2025cellflow}. These models typically learn an end-to-end mapping from a given perturbation to a corresponding gene expression profile, trained on large-scale perturbation-response datasets. However, prevailing approaches suffer from two critical limitations. First, they are heavily reliant on the scale, quality, and coverage of the training data. The data-hungry nature makes these models expensive to train. More importantly, it also limits their ability to generalize to novel perturbations that are not present in the training data~\citep{li2024systematic,ahlmann2024deep}. Second, these models operate as black boxes. While they may yield predictive outputs, they fail to provide clear, verifiable mechanistic explanations for predictions~\citep{hassija2024interpreting,noutahi2025virtual}. This lack of interpretability severely undermines their trustworthiness and utility in scientific discovery, making them difficult for biologists to rely upon for designing downstream experiments.

We argue that an ideal virtual cell model should not only provide accurate predictions but also be data-efficient, interpretable, and aligned with established biological principles. Rather than relying solely on statistical correlations, it ought to integrate fundamental biological knowledge to capture and anticipate cellular responses. Moreover, its reasoning process should be transparent, explicitly revealing the causal mechanisms that underlie predictions and grounding them in the well-established frameworks of cell signaling and gene regulation.

To overcome the above challenges, we introduce VCWorld, a cell-level white-box simulator(Figure \ref{fig:pipeline}). The core of VCWorld is a biological world model that simulates the dynamic response of a cell to drug perturbations. Instead of relying solely on statistical patterns, VCWorld integrates structured biological knowledge, such as signaling pathways, protein-protein interactions, and gene regulatory networks, with the iterative reasoning capabilities of Large Language Models (LLMs)~\citep{dubey2024llama,guo2025deepseek}. This design allows the model to generalize from limited training data by leveraging a vast repository of open-world biological knowledge. Crucially, VCWorld generates a transparent, traceable reasoning path for each prediction, offering a step-by-step mechanistic explanation that culminates in verifiable hypotheses. Furthermore, to facilitate finer-grained modeling, we introduce GeneTAK, a new benchmark derived from the large-scale Tahoe-100M dataset~\citep{zhang2025tahoe}. GeneTAK reframes cell-drug observations into gene-centric perturbation responses, mitigating data sparsity and enabling models to focus directly on the nuanced impact of a drug on individual genes. We highlight three main contributions of our study:
\begin{itemize}
\item We propose VCWorld, a novel cell-level white-box simulator architected as a biological world model. It combines structured biological knowledge with LLM-based reasoning, demonstrating a superior balance of data efficiency, interpretability, and predictive accuracy that overcomes the limitations of existing black-box models.
\item We construct and introduce GeneTAK, a new benchmark that transforms cell-drug perturbation data into single-gene response profiles. This allows for more granular and robust modeling of drug effects.
\item  We demonstrate that VCWorld achieves state-of-the-art performance on both the differential expression (DE) and directional (DIR) prediction tasks on the GeneTAK benchmark, validating the effectiveness of our approach.
\end{itemize}

\section{Related Work}

\subsection{Virtual Cell}
Virtual cell was originally a computational tool for simulating intracellular biochemical reactions, diffusion, membrane transport, and electro physiological processes~\citep{loew2001virtual}. With the development of AI, the concept has evolved into predictive  cell models that integrate large-scale multi-modal biological data to forecast, explain, and guide experimental hypotheses~\citep{bunne2024build,noutahi2025virtual}. Existing research can be broadly categorized into three approaches: data-driven methods combined with prior knowledge, such as GenePT~\citep{chen2024genept} and GEARS~\citep{roohani2024predicting}; large-scale pretrained foundational models, like scFoundation~\citep{hao2024large} and scGPT~\citep{cui2024scgpt}, trained on tens of millions of cells to learn general representations and excel in perturbation prediction; and generative modeling approaches, including CPA~\citep{lotfollahi2023predicting}, STATE~\citep{adduri2025predicting}, and CellFlow~\citep{klein2025cellflow}, which capture complex perturbation effects through decoupled latent spaces, state transitions, or flow-based generation. Further efforts, such as CellForge~\citep{tang2025cellforge}, aim to automate data analysis, literature review, and model design, pushing virtual cell toward self-directed evolution. 

Unlike previous approaches that mainly focus on end-to-end black-box models for cell perturbation prediction, our approach leverages a language model with reasoning and retrieval capabilities to construct a framework that emphasizes interpretability while also achieving superior performance.

\subsection{LLMs for Biological Reasoning and Prediction}

Recent research focus on leveraging Large Language Models (LLMs) for interpretable biological reasoning, aiming to move beyond traditional black-box predictions. A core approach in this area is the application of Chain-of-Thought (CoT) prompting~\citep{wei2022chain} to integrate multi-source heterogeneous data. For instance, frameworks like CoTox~\citep{park2025cotox} and DrugReasoner~\citep{ghaffarzadeh2025drugreasoner} combine chemical structures, biological pathways, and Gene Ontology (GO) terms to generate interpretable predictions for drug toxicity and approval processes. This paradigm has also been extended to fundamental biological contexts, such as the SUMMER framework~\citep{wu2025contextualizing}, which utilizes retrieval-augmented generation (RAG) to predict perturbation experiments under gene edits. To further enhance the reasoning capabilities of these models, researchers are actively exploring more advanced training methodologies. These studies~\citep{istrate2025rbio1,hasanaj2025multimodal} collectively demonstrate the strong potential of CoT to effectively integrate multimodal biological data within the virtual cell domain, enabling more nuanced and interpretable predictions compared to foundational models.

\subsection{Pert-seq Data Representation}  

Large-scale perturbation datasets~\citep{wu2024perturbench} are foundational for building predictive models of cellular behavior. Recently, scientists have gradually expanded the scope of perturbations. \cite{replogle2022mapping} enabled predictions of genetic perturbation effects on core cellular functions. \cite{nadig2024transcriptome} first introduced the dose effect in the perturbation. \cite{jiang2025systematic} aided inference of pathway signatures for perturbation-driven changes in disease scenarios. More recently, \cite{wu2025contextualizing} released PerturBase, a dedicated database for single-cell perturbation sequencing data, integrating 122 datasets from 46 studies across genetic and chemical perturbations.  A landmark in chemical perturbations is Tahoe-100M~\citep{zhang2025tahoe}, empowering AI models to predict context-dependent responses to small-molecule drugs across vast cellular diversity. Given the large size and coverage of the Tahoe-100M dataset, we use it as the primary dataset.

\section{Method}

\begin{figure}[t]
    \centering
    \includegraphics[width=1\linewidth]{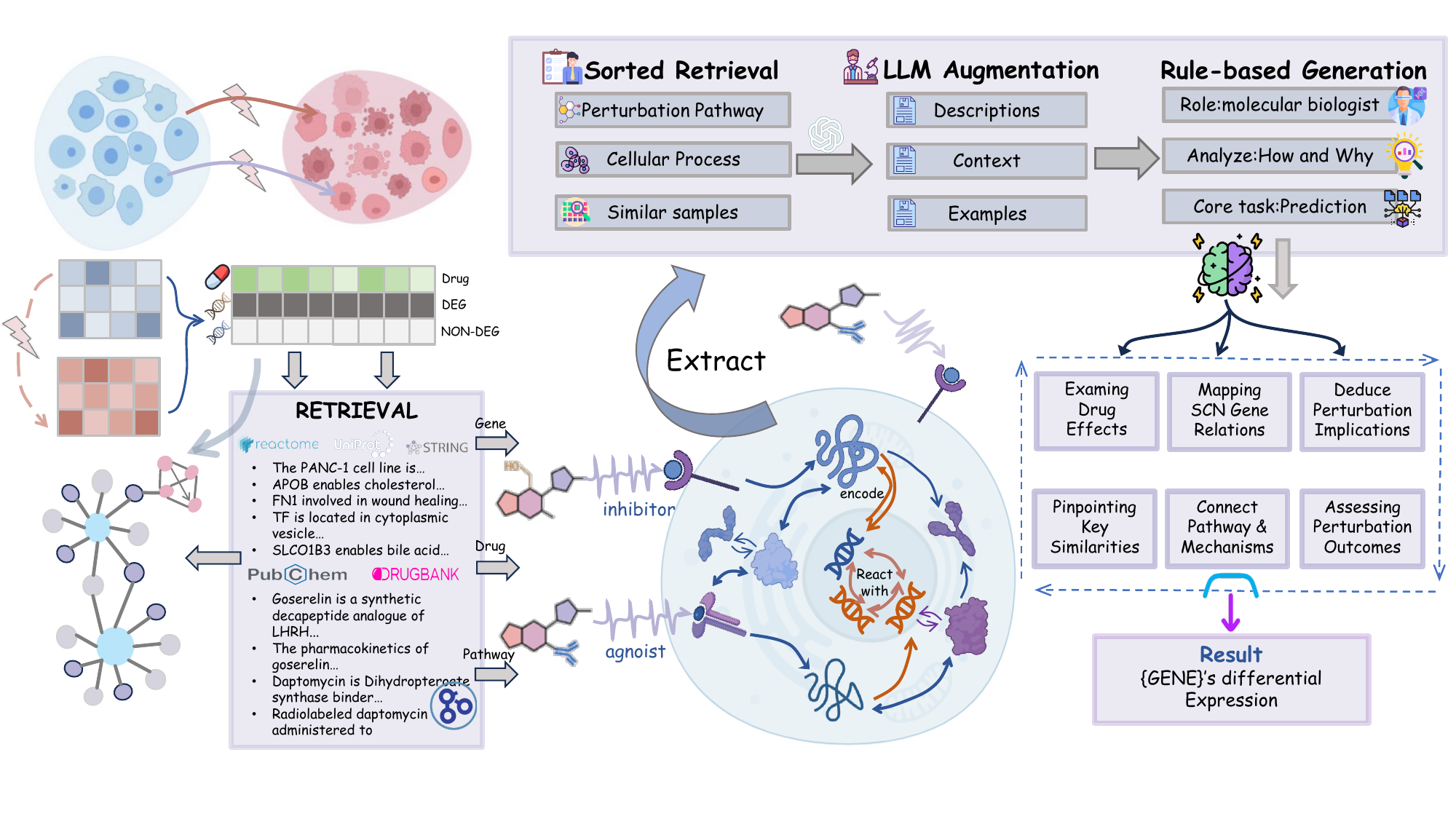}
    \vspace{-4mm}
    \caption{The pipeline of VCWorld. For a given query, VCWorld first retrieves multi-modal biological context from its integrated knowledge base, including pathway information, drug properties, and similar experimental samples. This context is then used to prompt a LLM, which is used to analyze mechanisms and infer how a specific gene will respond. The final output is a prediction for tasks such as differential expression and directional change.}
    \label{fig:pipeline}
    \vspace{-4mm}
\end{figure}

\subsection{Task Formulation}

The task of predicting single-cell gene expression responses to perturbations is conventionally defined as a regression problem. Given a dataset $D = \{(x_i, p_i, y_i)\}_{i=1}^N$, where $x_i, y_i \in \mathbb{R}^d$ are the pre- and post-perturbation expression profiles and $p_i$ is the perturbation, the goal is to learn a mapping function $f: \mathbb{R}^d \times \mathcal{P} \to \mathbb{R}^d$. However, the high dimensionality and sparsity of the data pose significant challenges to achieving accurate predictions at the individual gene level. To this end, we reformulate the problem as a \textit{gene-centric classification task}. The fundamental predictive unit is a triplet $(c, p, g)$, which queries the response of a specific gene $g$ to a perturbation $p$ in a given cell lineage $c$. For each such triplet, we derive a corresponding binary label, let's denote it as $l \in \{0, 1\}$, which represents the outcome for our two fine-grained classification tasks: \textnormal{(i)} \textbf{Differential Expression (DE):} $l=1$ if gene $g$ is differentially expressed, and $l=0$ otherwise. \textnormal{(ii)} \textbf{Directional Change (DIR):} $l=1$ for upregulation and $l=0$ for downregulation. This transformation converts the original dataset $D$ into a new, larger corpus of labeled triplets suitable for training a classifier.

These tasks can be addressed by a new paradigm that shifts from numerical regression to knowledge-based inference. The approach utilizes a reasoning engine, such as a Large Language Model (LLM) $L$, to process not only the query but also a rich set of biological context (BioContext). This context is dynamically retrieved from external knowledge bases (\textit{e.g.}, signaling pathways, protein-protein interactions). The problem is then transformed into a text-based inference function as follows:
$
f: L(c, p, g, \texttt{BioContext}) \to \texttt{Prediction}.
$

\begin{figure}[t]
    \centering
    \includegraphics[width=\linewidth]{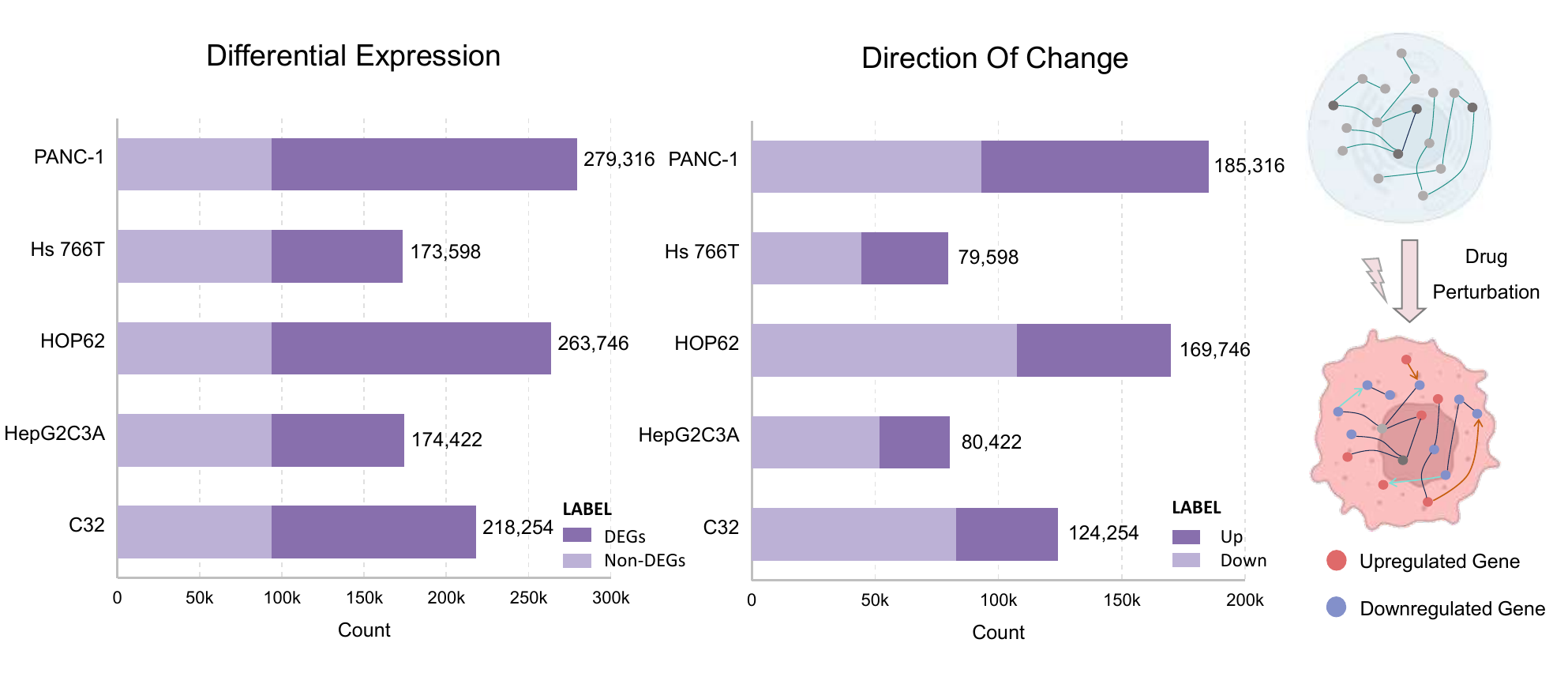}
    
    \caption{(a) Number of single-cell profiles per cell line. (b) Label distributions for the Differential Expression (DE) and Directional Change (DIR) tasks. (c) Schematic for identifying highly variable genes by comparing perturbed cells against a control group.}
    \label{fig2}
    \vspace{-4mm}
\end{figure}

\subsection{GeneTAK}

To facilitate the comprehensive and fair assessment of large language models for perturbation prediction, we present GeneTAK, a novel benchmark dataset. GeneTAK is curated from the Tahoe-100M raw expression matrix and comprises 5 distinct cell lines across 348 drug perturbations, designed explicitly to evaluate the generalization capabilities of models in this domain.
\paragraph{Dataset Curation.}

The GeneTAK expression matrix was first restricted to the 2,000 most highly variable genes, aligning it with the common input dimensions of generative perturbation prediction models. Subsequently, to ensure a diverse and representative selection of cell lines, we performed a Principal Component Analysis (PCA) visualization~\citep{greenacre2022principal} (see Appendix A). Guided by this analysis, we selected a final set of five cell lines that includes both a main cluster and several distinct outliers: C32, HOP62, HepG2C3A, Hs 766T, and PANC-1. Notably, this selection is consistent with the cell lines used in the work of STATE~\citep{adduri2025predicting}, ensuring a degree of comparability and fairness with previous benchmarks.

Then we generated labels for perturbation-gene pairs, denoted as $(p, g)$. The labels were determined by identifying differentially expressed genes (DEGs) for each perturbation using the Wilcoxon signed-rank test~\citep{woolson2007wilcoxon}. To ensure label quality, we applied strict criteria based on a significant adjusted p-value and consistent expression changes across biological replicates. The resulting dataset of labeled pairs was then split by perturbation into training and test sets at a 3:7 ratio. This specific ratio was intentionally chosen to simulate a challenging few-shot learning scenario, consistent with the evaluation framework used by~\citet{adduri2025predicting} and allowing for a direct comparison of model performance under low-data conditions. The split maintains a similar distribution of DEGs per perturbation in both sets. The final label distribution is shown in Figure \ref{fig2}, with a complete description of all data processing methods provided in Appendix \ref{appA}.

\subsection{VCWorld}

Our proposed framework, VCWorld (Figure \ref{fig:pipeline}), transforms biological prediction into a multi-stage reasoning process powered by a Large Language Model (LLM). It consists of three key stages: (1) generating rich, symbolic representations for all biological entities; (2) retrieving a support set of causal evidence guided by open-world biological knowledge; and (3) synthesizing all information through a Chain-of-Thought process to produce an interpretable prediction.

\subsection{Construction of the Open-World Biological Knowledge Graph}

To construct the open-world biological knowledge graph, we integrated several authoritative databases: PubChem, which provides large-scale compound structures and bioactivity data~\citep{kim2023pubchem}; DrugBank, which includes chemical, pharmacological, and clinical annotations of drugs and their target associations~\citep{knox2024drugbank}; UniProt, which offers comprehensive protein sequences and functional annotations~\citep{uniprot2019uniprot}; Gene Ontology (GO), which defines a unified ontology across molecular function, biological process, and cellular component~\citep{ashburner2000gene}; Reactome, which curates systematically organized molecular reactions and pathways~\citep{fabregat2018reactome}; STRING, which compiles protein–protein interactions from experiments and predictions~\citep{szklarczyk2021string}; and CORUM, which catalogs experimentally validated mammalian protein complexes~\citep{ruepp2007corum}. Together, these databases cover multi-scale biological knowledge from compounds and drugs to genes, proteins, pathways, and complexes, ensuring the comprehensiveness of the graph.

In the knowledge graph construction process, compounds, drugs, genes, proteins, pathways, and complexes were represented as entity nodes, while interactions, annotations, and hierarchical relationships were encoded as heterogeneous edges. Cross-database integration was achieved through standardized identifiers (\textit{e.g.}, InChIKey, UniProt ID, GO terms), followed by redundancy removal and conflict resolution. The resulting graph systematically represents biomolecules and their interactions, providing a foundation for cellular perturbation prediction.

\subsubsection{Generative Node Feature Representation via LLMs}
Let the open-world biological knowledge be denoted as $G = (\mathcal{V}, \mathcal{E}, \mathcal{R})$, where $\mathcal{V}$ is the set of biological entities (nodes), $\mathcal{R}$ is the set of relation types, and $\mathcal{E} \subseteq \mathcal{V} \times \mathcal{R} \times \mathcal{V}$ is the set of triples or edges. The first stage of our framework focuses on defining each biological entity $v \in \mathcal{V}$. To move beyond simple numerical vectors, we generate a rich symbolic representation for each node. For each node $v$, we first extract its local neighborhood subgraph, $N_k(v)$, defined as the set of all triples within a $k$-hop distance from $v$. We then construct a structured prompt $P_v$, using a template function $f_{\text{prompt}}$:
\begin{equation}
    \label{eq:prompt_gen}
    P_v = f_{\text{prompt}}(v, N_k(v)).
\end{equation}
This function serializes the node's core attributes and its neighborhood triples into a natural language query. Subsequently, our framework employs an LLM to function as a feature generator. The LLM $L$ then processes this prompt to generate a comprehensive textual description(see Appendix \ref{appP}), $d_v$, serving as the node’s initial feature representation:
\begin{equation}
    \label{eq:feature_gen}
    d_v = L (P_v)
\end{equation}
This process yields a context-aware representation $d_v \in \mathcal{T}$ (where $\mathcal{T}$ is the space of all possible texts) that preserves biological semantics, offering a more expressive alternative to static embeddings.

\subsubsection{Graph-guided causal evidence framework}

With rich node representations established, the second stage retrieves relevant experimental cases to form a basis for reasoning. Our training corpus $\mathcal{D}$, consists of $M$ labeled instances derived from the original data, can be formulated as $\mathcal{D} = \{(q_i, l_i)\}_{i=1}^M$. Here, each query $q_i$ is a triplet $(c_i, p_i, g_i)$, and $l_i$ is its associated ground-truth binary label ($l_i \in \{0, 1\}$). Given a new query $q_{input} = (c_\text{input}, p_\text{input}, g_\text{input})$, our goal is to construct an evidence support set $S(q_\text{input}) \subset \mathcal{D}$.

To ground the LLM's prediction in empirical data, we introduce a structured retrieval mechanism that goes beyond standard Retrieval-Augmented Generation (RAG)~\citep{lewis2020retrieval}. The core of this mechanism is to identify and retrieve the most relevant historical cases from our training corpus $\mathcal{D}$ for given input query $q_{\text{input}}$. To quantify this relevance, we score the similarity between the input query $q_{\text{input}}$ and each query $q_i \in \mathcal{D} $. Instead of relying solely on semantic similarity, our method computes a hybrid similarity score, $\text{Sim}(q_{\text{input}}, q_i)$, that also leverages knowledge-graph topology information:
\begin{equation}
    \label{eq:hybrid_sim}
    \text{Sim}(q_{\text{input}}, q_i) = \alpha \cdot \text{Sim}_{\text{sem}}(d_{q_{\text{input}}}, d_{q_i}) + (1-\alpha) \cdot \text{Sim}_{\text{struct}}(q_{\text{input}}, q_i)
\end{equation}
where $\text{Sim}_{\text{sem}}(\cdot)$ is the semantic similarity, calculated as the cosine similarity between the LLM-generated feature descriptions (from Eq.~\ref{eq:feature_gen}), and $\text{Sim}_{\text{struct}}(\cdot)$ is a graph-based structural similarity metric (\textit{i.e.}, path-based similarity). The hyperparameter $\alpha \in [0, 1]$ balances the contribution of the semantic and structural components. 

This score allows us to assemble a multifaceted evidence set. Instead of retrieving a single list of similar items, we retrieve two disjoint subsets by searching within predefined outcome groups as follows:

\textbf{Analogue Cases ($S_{\text{analog}}$):} The top-$k_a$ instances from the subset of $\mathcal{D}$ with a positive outcome (\textit{i.e.}, label $l=1$), ranked by their similarity to the input query.
    \begin{equation}
        \label{eq:analog_cases}
        S_{\text{analog}}(q_{\text{input}}) = \underset{q_i \in \{q \in \mathcal{D} \,|\, l=1\}}{\text{arg top-}k_a} \text{Sim}(q_{\text{input}}, q_i)
    \end{equation}

\textbf{Contrast Cases ($S_{\text{contrast}}$):} The top-$k_c$ instances from the subset of $\mathcal{D}$ with a negative outcome (i.e., label $l=0$), similarly ranked by similarity.
    \begin{equation}
        \label{eq:contrast_cases}
        S_{\text{contrast}}(q_{\text{input}}) = \underset{q_i \in \{q \in \mathcal{D} \,|\, l=0\}}{\text{arg top-}k_c} \text{Sim}(q_{\text{input}}, q_i)
    \end{equation}

The final evidence support set is the union: $S(q_{\text{input}}) = S_{\text{analog}} \cup S_{\text{contrast}}$. This curated set provides a balanced, contextual foundation upon which the final reasoning process is built.

\subsubsection{Evidence Synthesis Chain-of-Thought Reasoning}
In the final stage, the LLM acts as a computational biologist, which we formulate as a Chain-of-Thought (CoT) reasoning task. A final prompt, $P_{\text{CoT}}$, is synthesized from the symbolic representation of the query, $d_{q_{\text{input}}}$, and the retrieved evidence set, $S(q_{\text{input}})$:
\begin{equation}
    \label{eq:cot_prompt}
    P_{\text{CoT}} = f_{\text{CoT\_prompt}}(d_{q_{\text{input}}}, S_{\text{analog}}, S_{\text{contrast}})
\end{equation}
The function $f_{\text{CoT\_prompt}}$ formats these inputs using a structured template that first presents the biological query, followed by the lists of analogue and comparison cases as evidence. Finally, it instructs the LLM to provide step-by-step reasoning. (A detailed example is in Appendix \ref{appP}).

The LLM then processes this prompt to generate a text string, $O_{\text{final}}$, we then use a parsing function, $f_{\text{parse}}$, to extract the structured prediction $\hat{l}$ and the textual explanation $E$ from the output:
\begin{equation}
    \label{eq:cot_gen}
    O_{\text{final}} = L (P_{\text{CoT}})
\end{equation}
\begin{equation}
    \label{eq:parsing}
    (\hat{l}, E) = f_{\text{parse}}(O_{\text{final}})
\end{equation}
This process compels the LLM to explicitly integrate qualitative knowledge ($d_{q_{\text{input}}}$) with empirical evidence ($S(q_{\text{input}})$), producing a self-validating and fully interpretable output. In this work, we use Gemeni2.5-Flash~\citep{comanici2025gemini} as our reasoning model.

\section{Experiment}
\label{exp}
 \begin{table}[htbp]
    \centering
    \caption{Overall accuracy on DE and DIR tasks. The best results are shown in bold, and the second-best results are shown with underlines.}
    \label{accuracy}

    \resizebox{\textwidth}{!}{
        \begin{tabular}{llccccc}
            
            \toprule
            \textbf{Task} & \textbf{Model} & \textbf{C32} & \textbf{HepG2C3A} & \textbf{HOP62} & \textbf{Hs 766T} & \textbf{PANC-1} \\
            \midrule
            \multirow{13}{*}{\textbf{DE}} 
            & RANDOM & $0.50 \pm .00$ & $0.50 \pm .00$ & $0.50 \pm .00$ & $0.50 \pm .00$ & $0.50 \pm .00$ \\
            \cmidrule(lr){2-7}
            & GAT & $0.62 \pm .03$ & $0.54 \pm .02$ & $0.64 \pm .04$ & $0.57 \pm .01$ & $0.58 \pm .01$ \\
            & CPA & $0.17 \pm .04$ & $0.18 \pm .02$ & $0.21 \pm .03$ & $0.30 \pm .04$ & $0.17 \pm .02$ \\
            & scVI & $\underline{0.66} \pm .02$ & $0.48 \pm .01$ & $0.64 \pm .02$ & $0.61 \pm .01$ & $\underline{0.68} \pm .02$ \\
            & STATE & $0.17 \pm .02$ & $0.38 \pm .01$ & $0.41 \pm .01$ & $0.09 \pm .02$ & $0.47 \pm .01$ \\
            \cmidrule(lr){2-7}
            & \textbf{VCWorld} (Retrieval-only) & $0.57 \pm .01$ & $0.53 \pm .00$ & $0.58 \pm .01$ & $0.57 \pm .02$ & $0.58 \pm .01$ \\
            & \textbf{VCWorld} (w/o Biocontext) & $0.52 \pm .03$ & $0.54 \pm .03$ & $0.51 \pm .02$ & $0.52 \pm .01$ & $0.50 \pm .03$ \\
            & \textbf{VCWorld} (w/o CoT) & $0.61 \pm .03$ & $0.57 \pm .00$ & $0.59 \pm .01$ & $0.60 \pm .02$ & $0.56 \pm .03$ \\
            \cmidrule(lr){2-7}
            & \textbf{VCWorld} (Llama3-8B) & $0.35 \pm .01$ & $0.36 \pm .00$ & $0.41 \pm .01$ & $0.36 \pm .00$ & $0.39 \pm .02$ \\
            & \textbf{VCWorld} (Qwen3-4B) & $0.41 \pm .00$ & $0.47 \pm .01$ & $0.46 \pm .00$ & $0.49 \pm .02$ & $0.46 \pm .01$ \\
            & \textbf{VCWorld} (Qwen2.5-7B) & $0.54 \pm .01$ & $0.54 \pm .00$ & $0.55 \pm .00$ & $0.58 \pm .01$ & $0.64 \pm .02$ \\
            & \textbf{VCWorld} (Qwen2.5-14B) & $0.65 \pm .02$ & $\underline{0.66} \pm .01$ & $\mathbf{0.72} \pm .01$ & $\underline{0.67} \pm .00$ & $\mathbf{0.76} \pm .01$ \\
            & \textbf{VCWorld} (Gemini-2.5-Flash) & $\mathbf{0.70} \pm .00$& $\mathbf{0.68} \pm .00$ & $\underline{0.71} \pm .02$ & $\mathbf{0.68} \pm .00$ & $0.61 \pm .02$ \\
            \midrule
            \multirow{13}{*}{\textbf{DIR}} 
            & RANDOM & $0.50 \pm .00$ & $0.50 \pm .00$ & $0.50 \pm .00$ & $0.50 \pm .00$ & $0.50 \pm .00$ \\
            \cmidrule(lr){2-7}
            & GAT & $0.58 \pm .03$ & $0.61 \pm .01$ & $0.54 \pm .02$ & $0.55 \pm .01$ & $0.52 \pm .01$ \\
            & CPA & $0.22 \pm .01$ & $0.19 \pm .03$ & $0.17 \pm .01$ & $0.21 \pm .02$ & $0.15 \pm .01$ \\
            & scVI & $0.47 \pm .01$ & $0.46 \pm .01$ & $0.41 \pm .03$ & $0.40 \pm .01$ & $0.38 \pm .02$ \\
            & STATE & $0.49 \pm .01$ & $0.51 \pm .03$ & $0.50 \pm .01$ & $0.55 \pm .02$ & $0.51 \pm .01$ \\
            \cmidrule(lr){2-7}
            & \textbf{VCWorld} (Retrieval-only) & $0.47 \pm .01$ & $0.47 \pm .02$ & $0.47 \pm .01$ & $0.50 \pm .00$ & $0.51 \pm .01$ \\
            & \textbf{VCWorld} (w/o Biocontext) & $0.37 \pm .00$ & $0.36 \pm .01$ & $0.36 \pm .01$ & $0.17 \pm .03$ & $0.34 \pm .02$ \\
            & \textbf{VCWorld} (w/o CoT) & $0.42 \pm .01$ & $0.45 \pm .00$ & $0.51 \pm .01$ & $0.47 \pm .01$ & $0.45 \pm .00$ \\
            \cmidrule(lr){2-7}
            & \textbf{VCWorld} (Llama3-8B) & $0.37 \pm .01$ & $0.34 \pm .02$ & $0.38 \pm .04$ & $0.40 \pm .01$ & $0.35 \pm .01$ \\
            & \textbf{VCWorld} (Qwen3-4B) & $0.54 \pm .01$ & $0.55 \pm .00$ & $0.52 \pm .00$ & $0.48 \pm .02$ & $0.50 \pm .01$ \\
            & \textbf{VCWorld} (Qwen2.5-7B) & $0.56 \pm .02$ & $0.56 \pm .01$ & $0.55 \pm .00$ & $0.51 \pm .00$ & $0.53 \pm .01$ \\
            & \textbf{VCWorld} (Qwen2.5-14B) & $\underline{0.67} \pm .01$ & $\underline{0.59} \pm .00$ & $\underline{0.65} \pm .01$ & $\mathbf{0.71} \pm .00$ & $\underline{0.66}\pm .01$ \\
            & \textbf{VCWorld} (Gemini-2.5-Flash) & $\mathbf{0.72} \pm 0.01$ & $\mathbf{0.68} \pm .00$ & $\mathbf{0.67} \pm .00$ & $\underline{0.66} \pm .00$ & $\mathbf{0.69} \pm .00$ \\
            \bottomrule
        \end{tabular}
        
    } 
\end{table}

\subsection{baseline}

We benchmarked performance against several baselines using GeneTAK, including RANDOM, which is a naive baseline that is assumed to have a statistically accurate prediction of $50\%$ for any binary classification task without any training or prior knowledge; three published deep learning models: scVI~\citep{lopez2018deep} is a conditional variational self-encoder that conditionally models perturbation effects in potential space, CPA~\citep{lotfollahi2023predicting} recognizes novel perturbations by learning disentangled and linearly combined latent embeddings, STATE~\citep{adduri2025predicting}, the current SOTA model that demonstrates the best overall performance in the drug perturbation prediction task.

\subsection{Metrics}
We evaluate model performance on two primary aspects: differentially expressed gene (DEG) prediction performance and LLM reasoning robustness. For a fair comparison, predictions from baseline models that output continuous expression values (\textit{e.g.}, STATE, scVI) are first converted into binary value. This is achieved by applying the Wilcoxon Signed-Rank Test~\citep{woolson2007wilcoxon} to the predicted profiles, mirroring the ground-truth label generation process. All metrics below are calculated on these standardized outputs.
\paragraph{Metric for DEG Prediction Performance.} We evaluate DEG prediction performance using four standard classification metrics: accuracy, precision, recall, and F1 score. Accuracy provides an overall measure of the proportion of correctly classified genes, but because most genes are not differentially expressed (resulting in substantial class imbalance), Accuracy alone can be misleading. To obtain a more informative assessment of performance on the positive class (DEGs), we therefore also report Precision, which reflects the ability to avoid false positives, and Recall, which reflects the ability to avoid false negatives. The F1 score, defined as the harmonic mean of Precision and Recall, provides a single, balanced summary of these two aspects. In addition, to avoid bias toward models trained with regression-style objectives, we also compute the AUROC and AUPRC.

\paragraph{Metric for LLM Reasoning Robustness.} We define Q-score to measure the robustness of the LLM's reasoning process. It is the frequency with which the model abstains from making a prediction for a given query (\textit{e.g.}, by stating it has insufficient information). A lower rate signifies higher robustness. The detailed formulas for all metrics are provided in Appendix \ref{app:metrics}.

\begin{table}[!t]
\small
\centering
\caption{Performance comparison of different models on DE and DIR tasks.}
\label{F1}
\resizebox{\textwidth}{!}{ 
    \begin{tabular}{@{} l l c c c c c c c @{}}
    \toprule
    \textbf{Task} & \textbf{Model} & \textbf{Metric} & \textbf{C32} & \textbf{HepG2C3A} & \textbf{HOP62} & \textbf{Hs 766T} & \textbf{PANC-1} & \textbf{Average} \\
    \midrule
    \multirow{12}{*}{DE} 
     & \multirow{3}{*}{GAT}     & Precision & 0.63 & 0.56 & 0.58 & 0.50 & 0.63 & 0.58 \\
     &                          & Recall    & 0.42 & 0.43 & 0.46 & 0.39 & 0.47 & 0.43 \\
     &                          & F1-Score  & \underline{0.50} & \underline{0.49} & \underline{0.51} & \underline{0.44} & \underline{0.54} & \underline{0.49} \\
     \cmidrule(lr){2-9}
     & \multirow{3}{*}{STATE}   & Precision & 0.12 & 0.10 & 0.18 & 0.10 & 0.19 & 0.13 \\
     &                          & Recall    & 0.15 & 0.32 & 0.36 & 0.09 & 0.37 & 0.26 \\
     &                          & F1-Score  & 0.14 & 0.14 & 0.24 & 0.10 & 0.25 & 0.17 \\
    \cmidrule(lr){2-9}
     & \multirow{3}{*}{CPA}     & Precision & 0.11 & 0.10 & 0.14 & 0.07 & 0.17 & 0.12 \\
     &                          & Recall    & 0.02 & 0.02 & 0.02 & 0.01 & 0.02 & 0.02 \\
     &                          & F1-Score  & 0.03 & 0.03 & 0.04 & 0.02 & 0.04 & 0.03 \\
    \cmidrule(lr){2-9}
     & \multirow{3}{*}{scVI}    & Precision & 0.13 & 0.09 & 0.18 & 0.08 & 0.19 & 0.13 \\
     &                          & Recall    & 0.48 & 0.39 & 0.52 & 0.38 & 0.51 & 0.46 \\
     &                          & F1-Score  & 0.21 & 0.14 & 0.27 & 0.14 & 0.28 & 0.21 \\
    \cmidrule(lr){2-9}
     & \multirow{3}{*}{VCWorld} & Precision & 0.60 & 0.64 & 0.61 & 0.57 & 0.52 & 0.59 \\
     &                          & Recall    & 0.70 & 0.68 & 0.71 & 0.68 & 0.61 & 0.68 \\
     &                          & F1-Score  & \textbf{0.65} & \textbf{0.66} & \textbf{0.66} & \textbf{0.62} & \textbf{0.56} & \textbf{0.63} \\
    \midrule
    \multirow{15}{*}{DIR} 
     & \multirow{3}{*}{GAT}     & Precision & 0.62 & 0.54 & 0.64 & 0.57 & 0.58 & 0.59 \\
     &                          & Recall    & 0.37 & 0.39 & 0.49 & 0.44 & 0.52 & 0.44 \\
     &                          & F1-Score  & \underline{0.46} & \underline{0.45} & \underline{0.56} & \underline{0.50} & \underline{0.55} & \underline{0.50} \\
    \cmidrule(lr){2-9}
     & \multirow{3}{*}{STATE}   & Precision & 0.14 & 0.18 & 0.17 & 0.09 & 0.10 & 0.14 \\
     &                          & Recall    & 0.18 & 0.35 & 0.26 & 0.16 & 0.25 & 0.24 \\
     &                          & F1-Score  & 0.16 & 0.24 & 0.21 & 0.12 & 0.15 & 0.18 \\
    \cmidrule(lr){2-9}
     & \multirow{3}{*}{CPA}     & Precision & 0.10 & 0.10 & 0.15 & 0.10 & 0.03 & 0.10 \\
     &                          & Recall    & 0.02 & 0.02 & 0.02 & 0.03 & 0.19 & 0.06 \\
     &                          & F1-Score  & 0.03 & 0.03 & 0.04 & 0.05 & 0.05 & 0.04 \\
    \cmidrule(lr){2-9}
     & \multirow{3}{*}{scVI}    & Precision & 0.08 & 0.08 & 0.10 & 0.11 & 0.09 & 0.09 \\
     &                          & Recall    & 0.49 & 0.46 & 0.46 & 0.51 & 0.46 & 0.48 \\
     &                          & F1-Score  & 0.14 & 0.14 & 0.17 & 0.18 & 0.15 & 0.16 \\
    \cmidrule(lr){2-9}
     & \multirow{3}{*}{VCWorld} & Precision & 0.61 & 0.54 & 0.58 & 0.58 & 0.60 & 0.58 \\
     &                          & Recall    & 0.72 & 0.68 & 0.67 & 0.66 & 0.69 & 0.68 \\
     &                          & F1-Score  & \textbf{0.66} & \textbf{0.60} & \textbf{0.62} & \textbf{0.62} & \textbf{0.64} & \textbf{0.63} \\
    \bottomrule
    \end{tabular}
}
\end{table}

\subsection{Gene-Level Performance}

We consider the ability to capture gene-level expression changes as the most fundamental evaluation criterion. To systematically assess this, the task is stratified into two components: Differential Expression (DE) prediction, which identifies whether a gene's expression changes significantly; and Directional Change (DIR) prediction, which determines whether the expression is up-regulated or down-regulated. We use Accuracy to evaluate the models' performance in correctly predicting the ground-truth labels. The results are detailed in Table~\ref{accuracy}.

\paragraph{Differential Expression Task Performance.}  As shown in Table~\ref{accuracy}, our model, VCWorld, achieves the highest overall accuracy, frequently exceeding 0.70 across multiple cell lines. Among the baseline models, performance is highly polarized. scVI and GAT demonstrate strong capabilities in detecting gene changes, showing competitive performance that even rivals some LLM variants. This indicates that these traditional models are effective at recognizing that a perturbation has occurred. Conversely, CPA and STATE perform poorly on this task, with accuracies falling far below the random baseline of 0.50, suggesting that these models fail to capture the basic perturbation patterns in this specific experimental setup, likely producing highly skewed or incorrect predictions.

\paragraph{Directional Change Task Performance.} The DIR task is inherently more challenging as it requires inferring the direction of expression changes. Most traditional baselines suffer a significant accuracy drop from the DE task, exposing their inability to determine regulatory directions. The only exception is GAT (0.52--0.61), indicating that modeling gene relationships via graphs provides some directional signals. Conversely, VCWorld achieves stable and superior accuracies (0.65--0.72). By leveraging the biological knowledge and reasoning of LLMs, VCWorld overcomes traditional limitations, accurately inferring not just that a gene is perturbed, but exactly how it is regulated.

\subsection{Cell-Level and Population-Level Analysis}

\begin{table}[t]
\small
    \centering
    \caption{Comparison of AUROC and AUPRC on Differential Expression (DE) prediction.}
    \label{AUROC/AUPRC}
    \resizebox{\textwidth}{!}{
        \begin{tabular}{lllccccc}
            \toprule
            Task & Metric & Model & C32 & HepG2C3A & HOP62 & Hs766T & PANC-1 \\
            \midrule
            \multirow{20}{*}{\textbf{DE}} 
            & \multirow{10}{*}{\textbf{AUROC}} 
            & RANDOM & 0.50 & 0.50 & 0.50 & 0.50 & 0.50 \\
            & & GAT & \textbf{0.65} & \textbf{0.59} & \textbf{0.69} & \textbf{0.62} & \textbf{0.68} \\
            & & CPA & 0.20 & 0.17 & 0.15 & 0.18 & 0.14 \\
            & & scVI & 0.61 & 0.47 & \underline{0.68} & \underline{0.60} & 0.54 \\
            & & STATE & \underline{0.62} & 0.49 & 0.63 & 0.57 & \underline{0.58} \\
            & & \textbf{VCWorld} (Llama3-8B) & 0.32 & 0.34 & 0.28 & 0.40 & 0.33 \\
            & & \textbf{VCWorld} (Qwen2.5-7B) & 0.45 & \underline{0.51} & 0.56 & 0.51 & 0.53 \\
            & & \textbf{VCWorld} (Qwen3-4B) & 0.35 & 0.42 & 0.35 & 0.34 & 0.36 \\
            & & \textbf{VCWorld} (Qwen2.5-14B) & 0.42 & 0.40 & 0.38 & 0.56 & 0.48 \\
            & & \textbf{VCWorld} (Gemini-2.5-Flash) & 0.51 & 0.42 & 0.47 & 0.46 & 0.52 \\
            \cmidrule(lr){2-8} \\ 

            & \multirow{10}{*}{\textbf{AUPRC}} 
            & RANDOM & 0.61 & 0.58 & 0.68 & 0.73 & 0.78 \\
            & & GAT & 0.43 & 0.38 & 0.47 & 0.42 & 0.46 \\
            & & CPA & 0.21 & 0.32 & 0.24 & 0.22 & 0.19 \\
            & & scVI & 0.42 & 0.48 & 0.52 & 0.56 & 0.65 \\
            & & STATE & 0.37 & 0.42 & 0.44 & 0.36 & 0.44 \\
            & & \textbf{VCWorld} (Llama3-8B) & 0.62 & 0.58 & 0.49 & 0.57 & 0.52 \\
            & & \textbf{VCWorld} (Qwen2.5-7B) & 0.80 & 0.60 & \underline{0.83} & \underline{0.82} & \textbf{0.88} \\
            & & \textbf{VCWorld} (Qwen3-4B) & 0.76 & 0.54 & 0.76 & 0.72 & 0.76 \\
            & & \textbf{VCWorld} (Qwen2.5-14B) & \textbf{0.85} & \underline{0.71} & 0.82 & \textbf{0.89} & \underline{0.87} \\
            & & \textbf{VCWorld} (Gemini-2.5-Flash) & \underline{0.83} & \textbf{0.72} & \textbf{0.84} & 0.81 & 0.84 \\
            \bottomrule
        \end{tabular}
    }
\end{table}

We believe that examining the collective response of gene populations offers a more accurate representation of cellular changes. To evaluate this, we quantified the number of DEGs predicted across the GeneTAK dataset (see Appendix~\ref{appB}).

The results show clear patterns. scVI predicts 3–5 times more DEGs than the ground truth. STATE is less stable, but on average predicts about twice as many DEGs. These results indicate that both models amplify perturbation effects and classify genes as DEGs too aggressively. This explains their high Accuracy but low Precision, Recall, and F1 Score, which limits their reliability. In contrast, CPA consistently underestimates DEG counts, reflecting a limited ability to capture perturbation effects.

To obtain a more balanced view, we further computed Precision, Recall, F1 score (Table \ref{F1}), and the threshold-free metrics AUROC , AUPRC (Table \ref{AUROC/AUPRC}). The results confirm the above observations: (\romannumeral 1) VCWorld achieves the best overall performance, with high and balanced precision (0.59) and recall (0.68), resulting in the highest F1 score (0.63); (\romannumeral 2) scVI attains relatively high recall (0.46) but low precision (0.13), yet still stands out as the strongest baseline; (\romannumeral 3) STATE shows mediocre performance across all metrics, with a low overall F1 score of only 0.17; and (\romannumeral 4) both STATE and scVI amplify perturbation effects, with scVI predicting several times more DEGs than the ground truth and STATE predicting about twice as many. The threshold-free evaluation further sharpens these conclusions: although baselines such as GAT and scVI reach moderate AUROC values (approximately 0.60–0.65), AUROC can be overly optimistic in highly imbalanced settings like gene perturbation, where non-DEGs vastly outnumber DEGs. In contrast, AUPRC, which focuses on the positive class (perturbed genes), provides a more decisive assessment; here, VCWorld clearly dominates, achieving an average AUPRC above 0.80 (\textit{e.g.}, 0.85 on C32), while the baselines typically remain in the 0.40–0.50 range.

These findings suggest that deep learning models often lack explicit constraints to regulate DEG counts. They fail to capture the true intensity and scope of perturbation effects. In contrast, the design of VCWorld directly addresses these limitations, which explains its superior performance.

\begin{figure}[!tbp]
    \centering
    \includegraphics[width=1\linewidth]{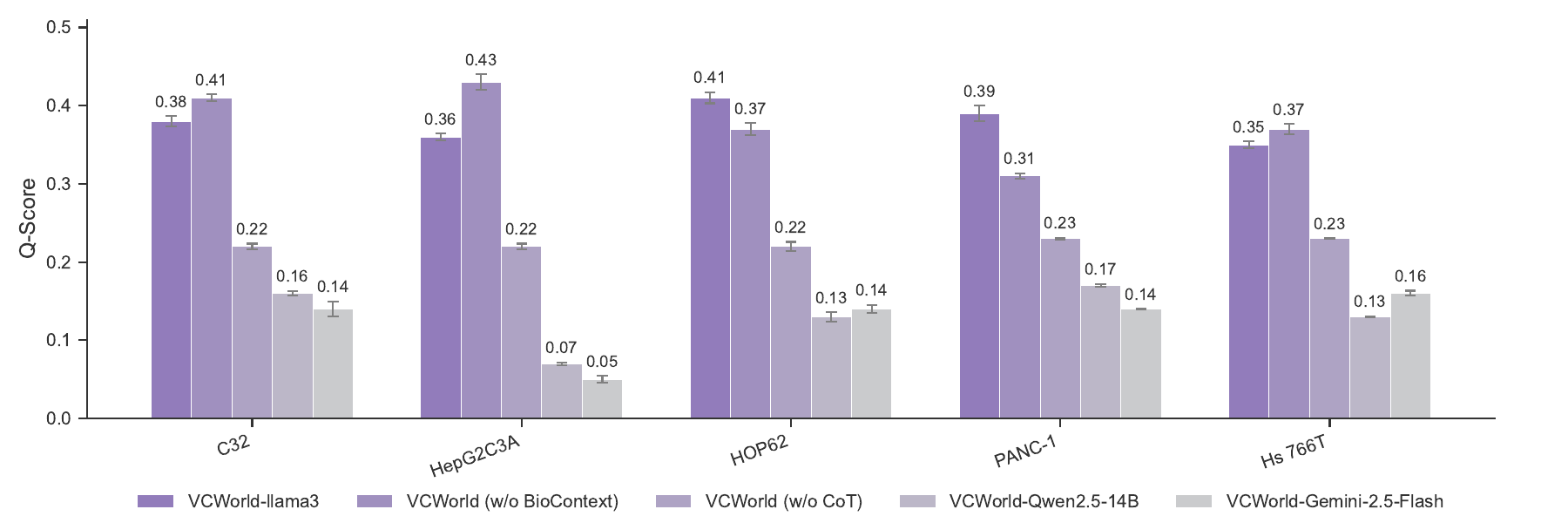}
    \caption{The ablation experiments of VCWorld.}
    \label{fig:ablation}
\end{figure}

\subsection{analysis and ablation study}

To evaluate the model’s tendency to avoid answering questions, we introduce \textbf{Q-score} as a dedicated metric for assessing LLM performance. Q-score is formally defined as a measure of the proportion of questions that are not answered normally, reflecting the model’s avoidance behavior when faced with complex or sensitive queries. Further details on the computation and usage of Q-score are provided in Appendix~\ref{app:qscore}. Figure \ref{fig:ablation} reports the Q-scores after ablating different components of the model. The ablation results reveal three key mechanisms driving VCWorld’s success:

\begin{itemize}
    \item \textbf{LLM Reasoning Capability is Critical:} Performance scales with the intelligence of the backbone model. We observe a clear trajectory from Llama3-8B (0.37) → Qwen2.5-14B (0.65) → Gemini-2.5-Flash (0.70) on the C32 cell line. This 84\% performance leap (Llama3-8B vs. Gemini-2.5-Flash) confirms that the task requires advanced reasoning, not just pattern recognition.
    \item \textbf{BioContext as a Cornerstone:} Removing biological context (VCWorld w/o BioContext) causes performance to plummet to near-random levels (0.51 average). This validates that the model relies on retrieved biological priors rather than hallucinating or guessing.
    \item \textbf{Chain-of-Thought (CoT) Standardization:} Enabling CoT improves the mean score by approximately 15\% (0.59 to 0.68). CoT acts as a regularizer, ensuring the model follows a logical inference path consistent with biological principles.
\end{itemize}

\section{Conclusion}
We demonstrate the efficacy of VCWorld in the task of predicting changes in gene expression following chemical perturbations. Our experiments show that VCWorld far outperforms current state-of-the-art black-box models. Its main contribution lies in aspects that are crucial for scientific utility. Through extensive ablation studies and case studies, we demonstrate that VCWorld is highly interpretable, with a human-readable, verifiable chain of reasoning for each prediction.

VCWorld still has room for improvement, with future directions including integrating a multi-agent framework to enhance autonomous reasoning and sample retrieval, extending generalization across diverse perturbation types such as gene, pathway, and combinatorial perturbations, and establishing more systematic benchmarks to comprehensively evaluate and compare specialized prediction models.

\newpage
\section*{Acknowledgments}
This study has been supported by the National Natural Science Foundation of China 62041209, Natural Science Foundation of Shanghai 24ZR1440600, the Science and Technology Commission of Shanghai Municipality 24510714300, the Lingang Lab Fund LGL-8888.

\normalem
\bibliography{iclr2026_conference}
\bibliographystyle{iclr2026_conference}

\newpage
\appendix

\begin{adjustwidth}{1cm}{}
\addcontentsline{toc}{section}{Appendix}
\startcontents[Appendix]
\printcontents[Appendix]{l}{1}{\setcounter{tocdepth}{2}}
\end{adjustwidth}

\section{GeneTAK}
\label{appA}

\subsection{Data Processing}

We normalized all gene counts to $\log({TP10k + 1})$ values (log-transformed UMI count per 10k), where the count $c_{ij}$ of gene $j$ in cell $i$ is mapped to:
$$\log \left( \frac{c_{ij}}{\sum_{j} c_{ij}} \cdot 10{,}000 + 1 \right).$$
To focus on the most variably expressed genes, the top 2,000 highly variable genes (HVGs) were selected using the \texttt{Seurat v3} Python package, and retained in the expression matrix for downstream analyses, with the HVG list exported for reproducibility. 
To determine differentially expressed genes (DEGs), we ran the Wilcoxon signed-rank test (Wilcoxon, 1945) with Benjamini-Hochberg correction (Benjamini \& Hochberg, 2000) between non-targeting control (NTC) cells and perturbed cells, for each perturbation.

We set $\text{FDR}_{\text{threshold}}$ and $\text{logFC}_{\text{threshold}}$ to screen perturbations with phenotypic effects, and $\text{PVAL}_{\text{threshold}}$ to screen negative samples (samples with no significant change in gene expression). Specifically, when 
$$\text{FDR}_{\text{threshold}} \leq 0.05 \quad \text{\&} \quad \text{logFC}_{\text{threshold}} \geq 0.25$$
(i.e., more than 1.28-fold change in expression), we consider them as samples with phenotypic effects. Based on this, when 
$$\text{PVAL}_{\text{neg\_threshold}} \geq 0.1,$$
we consider them as negative samples. In addition, we set $N_{\text{neg\_samples}} = 200$ to limit the number of negative samples, so as to avoid category imbalance.

\subsection{Distribution of GeneTAK}
\label{app:distri}
To keep consistent with the experimental settings in the STATE paper, the dataset of each cell line was split into training and testing sets at a ratio of 30:70.During the experiment, VCWorld only retrieved samples from the training set. Further details regarding the dataset and data split statistics are provided in Table~\ref{tab:3} and~\ref{tab:4}.

\begin{table}[h]
\small
\centering
\caption{Number of cells in each cell line.}
\begin{tabular}{lccc}
\toprule
Cell line & {Control} & {Perturbed} & {Sum} \\
\midrule
C32      & 29651  & 993514  & 1023165 \\
HepG2C3A & 20451  & 751387  & 771838 \\
HOP62    & 39683  & 1467690 & 1507373 \\
Hs 766T  & 17802  & 695932  & 713734 \\
PANC-1   & 43878  & 1534738 & 1578616 \\
\bottomrule
\end{tabular}
\label{tab:3}
\end{table}

\begin{table}[htbp]  
\small
\centering
\caption{GeneTAK statistics for differentially expressed genes.}
\label{tab:4}  
\begin{tabular}{llccccc}  
\toprule
Cell line & Split & {Total genes} & {Non-DE genes} & \multicolumn{3}{c}{\textbf{Differentially expressed genes}} \\  
\cmidrule(lr){5-7}  
& & & & {Total} & {Up-regulated} & {Down-regulated} \\  
\midrule
\multirow{2}{*}{C32}        & Train & 128293 & 55200 & 73093 & 24746 & 48344 \\
                             & Test  & 89961  & 38800 & 51161 & 16643 & 34518 \\
\multirow{2}{*}{HepG2C3A}  & Train & 102253 & 55200 & 47053 & 16820 & 30233 \\
                             & Test  & 72169  & 38800 & 33369 & 11629 & 21740 \\
\multirow{2}{*}{HOP62}     & Train & 154969 & 55200 & 99769 & 36615 & 63154 \\
                             & Test  & 108777 & 38800 & 69977 & 25472 & 44505 \\
\multirow{2}{*}{Hs 766T}   & Train & 101707 & 55200 & 46507 & 20774 & 25733 \\
                             & Test  & 71891  & 38800 & 33091 & 14420 & 18671 \\

\multirow{2}{*}{PANC-1}    & Train & 163748 & 55200 & 108548 & 54081 & 54467 \\
                             & Test  & 115568 & 38800 & 76768 & 38131 & 38637 \\  
\bottomrule
\end{tabular}
\end{table}

\section{Experiment Details}
\label{appB}
\subsection{Metrics}
\label{app:metrics}
\subsubsection{Accuracy}
Accuracy measures the proportion of correctly predicted samples among all samples. It is defined as:
\[
\text{Accuracy} = \frac{TP + TN}{TP + TN + FP + FN},
\]
where $TP$, $TN$, $FP$, and $FN$ represent the numbers of true positives, true negatives, false positives, and false negatives, respectively.

\paragraph{Definition of Confusion Matrix Components.}
Given binary ground-truth labels $y \in \{0,1\}$ and model predictions $\hat{y} \in \{0,1\}$:
\[
TP = \sum_{i=1}^{N} \mathbb{I}(y_i = 1 \land \hat{y}_i = 1),
\qquad
TN = \sum_{i=1}^{N} \mathbb{I}(y_i = 0 \land \hat{y}_i = 0),
\]
\[
FP = \sum_{i=1}^{N} \mathbb{I}(y_i = 0 \land \hat{y}_i = 1),
\qquad
FN = \sum_{i=1}^{N} \mathbb{I}(y_i = 1 \land \hat{y}_i = 0),
\]
where $\mathbb{I}(\cdot)$ is the indicator function.

\subsubsection{F1-score}
F1-score is the harmonic mean of Precision and Recall, balancing prediction correctness and coverage of true positives. The metrics are defined as:
\[
\text{Precision} = \frac{TP}{TP + FP}, 
\qquad
\text{Recall} = \frac{TP}{TP + FN},
\]
\[
\text{F1} = 2 \cdot \frac{\text{Precision} \cdot \text{Recall}}
{\text{Precision} + \text{Recall}}.
\]

\subsubsection{AUROC \& AUPRC}
The Area Under the Receiver Operating Characteristic Curve (AUROC) measures the model's ability to distinguish between classes. The True Positive Rate (TPR) and False Positive Rate (FPR) are defined as:
\[
\text{TPR} = \frac{TP}{TP + FN}, \qquad
\text{FPR} = \frac{FP}{FP + TN}.
\]
AUROC is computed as:
\[
\text{AUROC} = \int_{0}^{1} \text{TPR}(FPR) \, d(FPR).
\]

The Area Under the Precision–Recall Curve (AUPRC) quantifies the trade-off between Precision and Recall:
\[
\text{AUPRC} = \int_{0}^{1} \text{Precision}(Recall) \, d(Recall),
\]
where Precision and Recall are computed as defined above.

\subsubsection{Q-score}
\label{app:qscore}
\[
\text{Q-score} = 1-\frac{AP}{QP + AP}
\]
where $AP$ means the number of answered prompts, $QP$ means the number of prompts without a normal answer.

\subsection{Baselines}
\label{appT}
All baselines were trained on 30\% of the 45 cell lines from Tahoe-100m and 5 test cell lines, and all of our GeneTAK data was extracted from the remaining 70\%, which eliminates the possibility of data leakage, Table5 is the result obtained from the inference of all baseline models on the five cell lines.

\textbf{GAT.}  We moved beyond the disjoint subgraph design mentioned in the original manuscript. We integrated multiple knowledge sources (PubChem, DrugBank, GO, UniProt, Reactome) into a single connected graph structure, using standardized identifiers to align entities across databases. We used 70\% of the labeled GeneTAK instances for training the GAT model, with the prediction task being a three-class classification comprising up, down, and no change.The source code of training is available at \url{https://github.com/GENTEL-lab/VCWorld}.

\textbf{CPA} is a framework to learn the effects of perturbations at the single-cell level. CPA encodes and learns phenotypic drug responses across different cell types, doses, and combinations. The source code of CPA is available at \url{https://github.com/theislab/cpa}. In our experiments, we use the official implementation with the hyperparameters.

\textbf{scVI} is a family of probabilistic models that perform many analysis tasks across single-cell, multi-omics, and spatial omics data. The source code of scVI is available at \url{https://github.com/scverse/scvi-tools}. In our experiments, we adopt the standard scVI model with the hyperparameters in the official GitHub repository.

\textbf{STATE} is a machine learning model that predicts cellular perturbation responses across diverse contexts. The source code of STATE is available at \url{https://github.com/ArcInstitute/state}. In our experiments, we use the official implementation with a LLaMA-style transformer backbone and the key hyperparameters.

\subsection{Large Language Models}
\textbf{Llama-3.1} is Meta’s 8B-parameter instruction-tuned chat model optimized for multilingual dialogue, outperforming many open and closed models on standard benchmarks.The source code of Llama3.1-8B-Instruct is available at \url{https://huggingface.co/meta-llama/Llama-3.1-8B-Instruct}.

\textbf{Qwen2.5-7B-Instruct} is a 7.6B-parameter instruction-tuned open-source model with strong performance and efficiency on multilingual tasks, long-context understanding, and structured output generation.
The source code of Qwen2.5-7B-Instruct is available at \url{https://huggingface.co/Qwen/Qwen2.5-7B-Instruct}.

\textbf{Qwen3-4B-Instruct-2507} is the updated 4B-parameter Qwen3 instruction model with marked improvements in instruction following, reasoning, math, coding, and tool use, well-suited for lightweight multi-purpose deployment.The source code of Qwen3-4B-Instruct is available at \url{https://huggingface.co/Qwen/Qwen3-4B-Instruct-2507}.

\textbf{Qwen2.5-14B-Instruct} is a ~14B-parameter instruction-tuned model in the Qwen2.5 family, offering significantly improved coding and math reasoning capabilities over Qwen2 for more demanding applications.The source code of Qwen2.5-14B-Instruct is available at \url{https://huggingface.co/Qwen/Qwen2.5-14B-Instruct}.

\textbf{Gemini-2.5-Flash} is Google’s price-performance-optimized Gemini model with built-in “thinking” capabilities, designed for large-scale, low-latency, high-throughput and agentic workloads.

\subsection{Inference Parameters Settings}
For all models, we standardized the inference parameters with \texttt{temperature=0.6} and \texttt{top\_k = 0.9} to balance creativity and determinism.

\subsection{Cost and Latency Analysis} 
We monitored cost and latency using the OpenRouter API. To ensure a fair benchmark, the input prompt length was standardized to approximately 2,600 tokens for all models. The inference output tokens varied based on the reasoning length of each model. The detailed cost/latency breakdown for processing 1,000 samples is provided in Table \ref{tab:model_cost_latency}.

\begin{table}[htbp]
  \centering
  \caption{Model Cost and Latency(s)}
  \label{tab:model_cost_latency} 
  \resizebox{\textwidth}{!}{ 
      \begin{tabular}{lcccccc} 
        \toprule
        Model Name          & Input Price & Output Price & Input Token & Output Token & Latency(s) & Total Cost(\$) \\
        \midrule
        LLAMA3 8B           & 0.03        & 0.06         & 2600        & 1300         & 0.31       & 0.156          \\
        Qwen2.5-7B          & 0.04        & 0.1          & 2600        & 2700         & 0.42       & 0.374          \\
        Qwen2.5-14B         & 0.05        & 0.22         & 2600        & 2700         & 0.51       & 0.724          \\
        Qwen3-4B            & 0           & 0            & 2600        & 2700         & 0.68       & 0              \\
        Gemini-2.5-Flash     & 0.30        & 2.50         & 2600        & 1400         & 0.56       & 4.28           \\
        \bottomrule
      \end{tabular}
  }
  \end{table}

\begin{table}[H]
\centering
\small
\caption{Statistics of predicted number of DEGs}
\label{tab:de_task_results}
\begin{tabular}{lllccccc}
    \toprule
    \textbf{Task} & \textbf{Model} & \textbf{Count} & {\textbf{C32}} & {\textbf{HepG2C3A}} & {\textbf{HOP62}} & {\textbf{Hs\_766T}} & {\textbf{PANC-1}} \\
    \midrule

    \multirow{7}{*}{DE} & Truth & Number & 128293 & 102253 & 154969 & 101707 & 163748 \\
    \cmidrule{2-8}

    & \multirow{2}{*}{STATE} & Prediction & 157758 & 366238 & 317589 & 86684 & 316434 \\
    & & Correct & 19488 & 33122 & 55717 & 9077 & 60640 \\
    \cmidrule{2-8}

    & \multirow{2}{*}{CPA} & Prediction & 21475 & 21204 & 19524 & 19838 & 19253 \\
    & & Correct & 2424 & 2076 & 2755 & 1394 & 3178 \\
    \cmidrule{2-8}

    & \multirow{2}{*}{scVI} & Prediction & 460477 & 466868 & 449047 & 470606 & 435034 \\
    & & Correct & 60973 & 40108 & 79858 & 38969 & 83801 \\
    \bottomrule
\end{tabular}
\end{table}

\subsection{Evaluation Strategy}
Our evaluation strictly adheres to a zero-shot Drug Prediction protocol. Specifically:
\begin{itemize}
\item When predicting the effect of a test drug $p_{test}$, the retrieval corpus $\mathcal{D}$ consists exclusively of drugs from the training set.
\item The model never has access to the ground truth or the entity of $p_{test}$ itself during retrieval.
\item Inference relies solely on identifying functionally analogous cases (based on MoA and chemical structure) from the training set to extrapolate the potential effects of the unseen drug.
\end{itemize}

\subsection{Knowledge Graph-to-Text Pipeline for Prompt Generation}
\label{appendix:kg_to_prompt}

To ensure the factual grounding and mitigate the risk of hallucination in our language model, we developed a deterministic pipeline to programmatically construct input prompts from multiple structured biological knowledge graphs (KGs). This process transforms machine-readable graph data into a human-readable, factual context that guides the model's summarization task. The pipeline consists of three primary stages: data integration and pre-processing, rule-based templated verbalization, and fact aggregation for prompt assembly.

\subsubsection{Data Integration and Pre-processing}
Our pipeline integrates information from a diverse set of public biological knowledge bases, including Ensembl and UniProt for descriptive annotations, and graph-structured databases such as the Gene Ontology (GO), Reactome, CORUM, BioPlex, and STRING for relational information. Upon ingestion, a pre-processing step is applied to enhance data quality. Notably, we filter out high-degree nodes from relational graphs (\textit{e.g.}, generic GO terms like ``protein binding'' or highly promiscuous interactors in STRING). This step reduces noise and focuses the downstream context on more specific and informative biological relationships.

\subsubsection{Retrieval Strategy Analysis}

\paragraph{Sensitivity Analysis.} To validate this choice, we conducted an ablation study increasing the neighborhood depth from 1 to 3 (Table \ref{tab:accuracy_neighborhood_depth}). As shown in the additional results, increasing the depth significantly increased computational resource consumption and context length but resulted in negligible performance gains. This confirms that a 1-hop neighborhood is the optimal trade-off between efficiency and information density for this task.

\begin{table}[htbp]
  \centering
  \caption{Accuracy Performance On Neighborhood Depth Variations Experiment}
  \label{tab:accuracy_neighborhood_depth}
  \begin{tabular}{llccccc}
    \toprule
    Task       & Hop  & C32  & HepG2C3A & HOP62 \\
    \midrule
    \textbf{DE} & 1    & 0.70 & 0.68     & 0.71  \\
               & 2    & 0.72 & 0.65     & 0.70  \\
               & 3    & 0.69 & 0.67     & 0.70  \\
    \bottomrule
  \end{tabular}
\end{table}

\subsubsection{Decoupled Retrieval Strategy}

We employ a decoupled retrieval strategy for drugs and genes, utilizing a weighted combination of semantic and structural similarity.

\paragraph{Drug Retrieval ($\alpha=0.7$).}
We assign a weight of $\alpha=0.7$ to MoA-based semantic similarity and $1-\alpha=0.3$ to structure-based similarity (molecular fingerprints). This reflects the assumption that Mechanism of Action serves as the most direct reference for drug perturbation effects, while molecular structure provides complementary information related to scaffolds and functional groups.

\paragraph{Gene Retrieval ($\alpha=0.7$).}
We set the weight for functional semantic annotation to $\alpha=0.7$ and for PPI-based structural similarity to $1-\alpha=0.3$. Semantic annotations are prioritized due to their curated and generally high-confidence nature. Conversely, recent studies~\citep{zhang2019predicting} indicate that PPI networks may contain false-positive links. To mitigate noise, we assign a lower weight to the graph-derived component.

\subsubsection{Retrieval-Set Size}
We evaluate retrieval-set sizes of 5 and 10 (Table~\ref{tab:retrieval_set_response_rate}).  
The main effect lies in the model's \emph{willingness to answer}.  
When the size is reduced to 5, the definitive response rate on the C32 cell line drops from 87\% to 64\%, as the model more frequently abstains due to insufficient evidence, effectively reducing hallucination but also degrading overall performance.

\begin{table}[htbp]
  \centering
  \caption{Response Rate On Retrieval-set Size Experiment}
  \label{tab:retrieval_set_response_rate}
  \begin{tabular}{lcccccc}  
    \toprule
    Task & Retrieval & C32 & HepG2C3A & HOP62 & Average \\
    \midrule
    \multirow{2}{*}{\textbf{DE}} & 5  & 0.64 & 0.67 & 0.60 & 0.64 \\
                                & 10 & 0.87 & 0.87 & 0.88 & 0.87 \\
    \bottomrule
  \end{tabular}
\end{table}

\subsection{Rule-Based Templated Verbalization}
The core of our methodology is a rule-based verbalization engine. Recognizing that each knowledge source possesses a unique data schema, we designed a specific template to translate its structured entries into natural language declarative statements. 

For instance:
\begin{itemize}
\item A relationship from CORUM, representing \texttt{(Gene\_A, is\_member\_of, Complex\_X, in\_cell\_line\_Y)}, is verbalized as: \texttt{Gene A is a member of the Complex X in cell line Y}.
\item A protein-protein interaction from STRING, representing \texttt{(Gene\_A, interacts\_with, Gene\_B)} with associated evidence types \texttt{[experimental, database]}, is translated to: \texttt{Based on evidence from experimental evidence and database evidence, Gene A may physically interact with Gene B}.
\item Functional annotations from UniProt are extracted and formatted as statements like: \texttt{Functions of Gene A include: [function description]}.
\end{itemize}
This templated approach ensures that the conversion is consistent, deterministic, and preserves the semantic content of the original source.

\subsection{Fact Aggregation and Prompt Assembly}

For each target gene, we aggregate declarative statements (facts) collected from all integrated sources into a single set. To automatically remove duplicates arising from overlapping databases, we use a set data structure. To keep the prompt concise and reduce the cognitive load on the model, we limit the maximum number of facts included, giving priority to high-confidence sources such as UniProt and curated pathway databases.  

The resulting collection is then formatted as a bulleted list and inserted into a master prompt template under the heading ``\texttt{Domain Knowledge:}''. This final prompt conveys to the language model its role, a clearly defined task, operational constraints, and the essential, factually grounded context required to produce an accurate and reliable summary.

Given the imbalanced nature of the GeneTAK dataset, we agree that AUPRC provides a more informative assessment of performance on the minority class (differentially expressed genes).

\section{Case Study}

To qualitatively assess the explainability and reasoning capability of VCWorld, as well as its alignment with relevant wet-lab findings, we illustrate a representative case. The example focuses on the model’s prediction of the effect of Larotrectinib on the gene MKI67.

\textbf{Larotrectinib} is a precision-targeted therapy that inhibits TRK kinase activity, thereby blocking the proliferative and survival signaling pathways driven by NTRK fusions in tumors. \textbf{MKI67} encodes the nuclear protein \textbf{Ki-67}, a well-established marker of cellular proliferation. High expression of Ki-67 indicates active cell division and is frequently used in oncology to assess tumor proliferation, guide treatment selection, and predict prognosis. 

\begin{figure}[tbp]
    \centering
    \includegraphics[width=\linewidth]{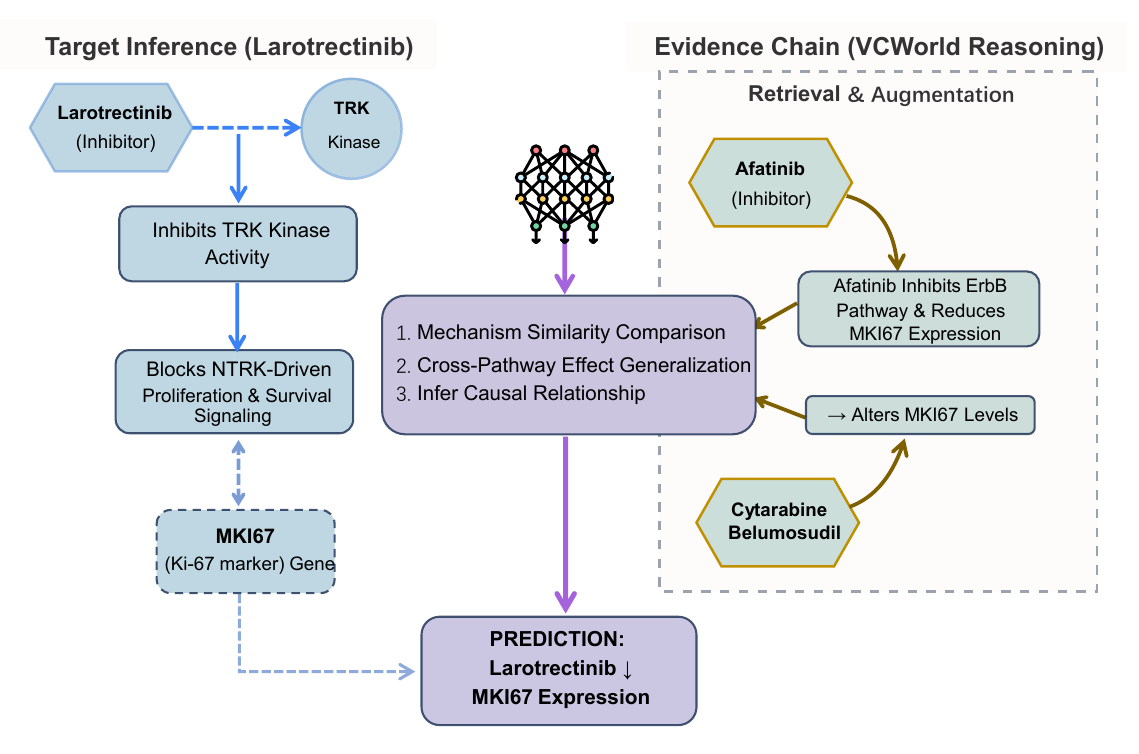}
    
    \caption{VCWorld Inference on (Larotrectinib,MKI67). The schematic illustrates the inference process for Larotrectinib-induced MKI67 downregulation, integrating direct target inference with retrieved biological evidence (BioContext). The LLM synthesizes these inputs via Chain-of-Thought (CoT) reasoning to predict causal relationships. }
    \label{case_study}
    \vspace{-4mm}
\end{figure}

Although no direct evidence has previously linked \textbf{Larotrectinib} with \textbf{MKI67} expression, \textbf{VCWorld} provided a plausible reasoning pathway that supports their association (Figure \ref{case_study}). The model first retrieved Afatinib---a drug with a similar mechanism---as a positive reference. Evidence showing that Afatinib suppresses the ErbB signaling pathway and reduces MKI67 expression suggested that inhibition of upstream kinases can influence proliferative markers. Furthermore, perturbations induced by Cytarabine and Belumosudil also led to alterations in \textbf{MKI67} levels, reinforcing its role as a reliable indicator of anti-proliferative effects. 

Based on this reasoning chain, VCWorld inferred that Larotrectinib downregulates MKI67 expression, a conclusion consistent with recent findings reporting that \textit{``treatment with larotrectinib led to a reduction in proliferating cells (5.73\% Ki67-positive cells)''}~\citep{schmid2024erk}. Consistently, immunofluorescence (IF) analysis revealed decreased Ki-67 signal intensity following Larotrectinib treatment~\citep{kong2022larotrectinib}.

Besides, we also show the original output results of the model used for reasoning about (Larotrectinib, MKI67), presented using the original markdown text.

\begin{Verbatim}
# Answer for Drug: Larotrectinib, Gene: MKI67

1) **Which of the observed perturbing drugs are most similar to Larotrectinib (if any, including Larotrectinib itself)?**
The most similar observed perturbing drug to Larotrectinib is **Afatinib (Example 9)**.
*   **Larotrectinib:** Is an inhibitor of tropomyosin receptor kinases (Trk) A, B, and C. It prevents neurotrophin-Trk interaction and subsequent Trk activation, inhibiting downstream signaling pathways that regulate cell growth, differentiation, and survival.
*   **Afatinib:** Is n irreversible ErbB family blocker that covalently binds to and inhibits the kinase domains of EGFR, HER2, and HER4. This inhibits tyrosine kinase autophosphorylation, downregulating ErbB signaling, and affecting genes involved in cell proliferation and survival.
Both drugs are kinase inhibitors that target receptor tyrosine kinases (Trk in Larotrectinib's case, ErbB in Afatinib's case) and primarily impact downstream signaling pathways related to cell proliferation and survival.

2) **When using Larotrectinib or similar drugs, what downstream pathways or genes are differentially expressed? Justify your answer with the observed outcomes.**
Based on the description of Larotrectinib, it is expected to suppress the expression of genes involved in **cell growth, differentiation, and survival** that are regulated by Trk signaling.
The most similar drug, **Afatinib (Example 9)**, which also inhibits receptor tyrosine kinases involved in cell proliferation and survival, resulted in **differential expression of MKI67**. This indicates that perturbations of receptor tyrosine kinase pathways affecting cell growth and proliferation indeed lead to changes in MKI67 expression.
Other drugs impacting MKI67 (e.g., Belumosudil affecting STAT3/5, Cytarabine causing DNA damage/cell cycle control) also point to pathways that regulate cell proliferation and division as being differentially expressed.


3) **What perturbations of upstream pathways or genes result in differential expression of MKI67 or similar genes (if any)? Justify your answer with the observed outcomes.**
MKI67 is a marker of cell proliferation and is impacted by cell cycle perturbations. The examples show that various perturbations targeting cell growth, cell cycle control, and cellular stress responses result in differential expression of MKI67:
*   **Inhibition of growth-related receptor tyrosine kinases:** **Afatinib** (ErbB family blocker, affects cell proliferation and survival, Example 9) resulted in differential expression of MKI67.
*   **Modulation of Rho-associated coiled-coil containing protein kinases (ROCK) and STAT pathways:** **Belumosudil** (ROCK2 inhibitor, modulates STAT3/STAT5, Example 2) resulted in differential expression of MKI67. STAT pathways are often involved in cell growth and differentiation.
*   **Induction of DNA damage and replication stress:** **Cytarabine** (DNA polymerase inhibitor, causes DNA damage and replication stress, affecting cell cycle control, Example 10) resulted in differential expression of MKI67.
*   **Induction of oxidative stress/cellular stress response:** **Verteporfin** (photosensitizer, generates reactive oxygen species, inducing oxidative stress and cellular stress response, Example 6) resulted in differential expression of MKI67.
*   **Activation of metabolic pathways:** **Rosiglitazone** (PPAR-gamma agonist, affects glucose metabolism, Example 7) resulted in differential expression of MKI67, suggesting a link between metabolism and proliferation.
Perturbations that interfere with fundamental processes like DNA replication, cell cycle progression, growth signaling, or induce significant cellular stress tend to impact MKI67 expression.

4) Is a perturbation of Larotrectinib in MKI67 likely to result in differential expression of MKI67? For example, if 2) and 4) are unrelated or only indirectly related, it is unlikely we will observe differential expression. On the other hand, if 2) and 4) significantly overlap in specific genes or pathways, we may observe differential expression. Your final answer should end with one of these three options and nothing else.
Yes. Perturbation of Larotrectinib results in differential expression of MKI67.
\end{Verbatim}

\section{Prompts}

We provide the prompts used by VCWorld and the model's response in the Appendix~\ref{appC}. It clearly demonstrates that VCWorld is interpretable, capable of aligning with real-world biological knowledge, and has a traceable thinking process and final result.

\label{appC}
\begin{figure}

    \centering
    \includegraphics[width=0.9\linewidth]{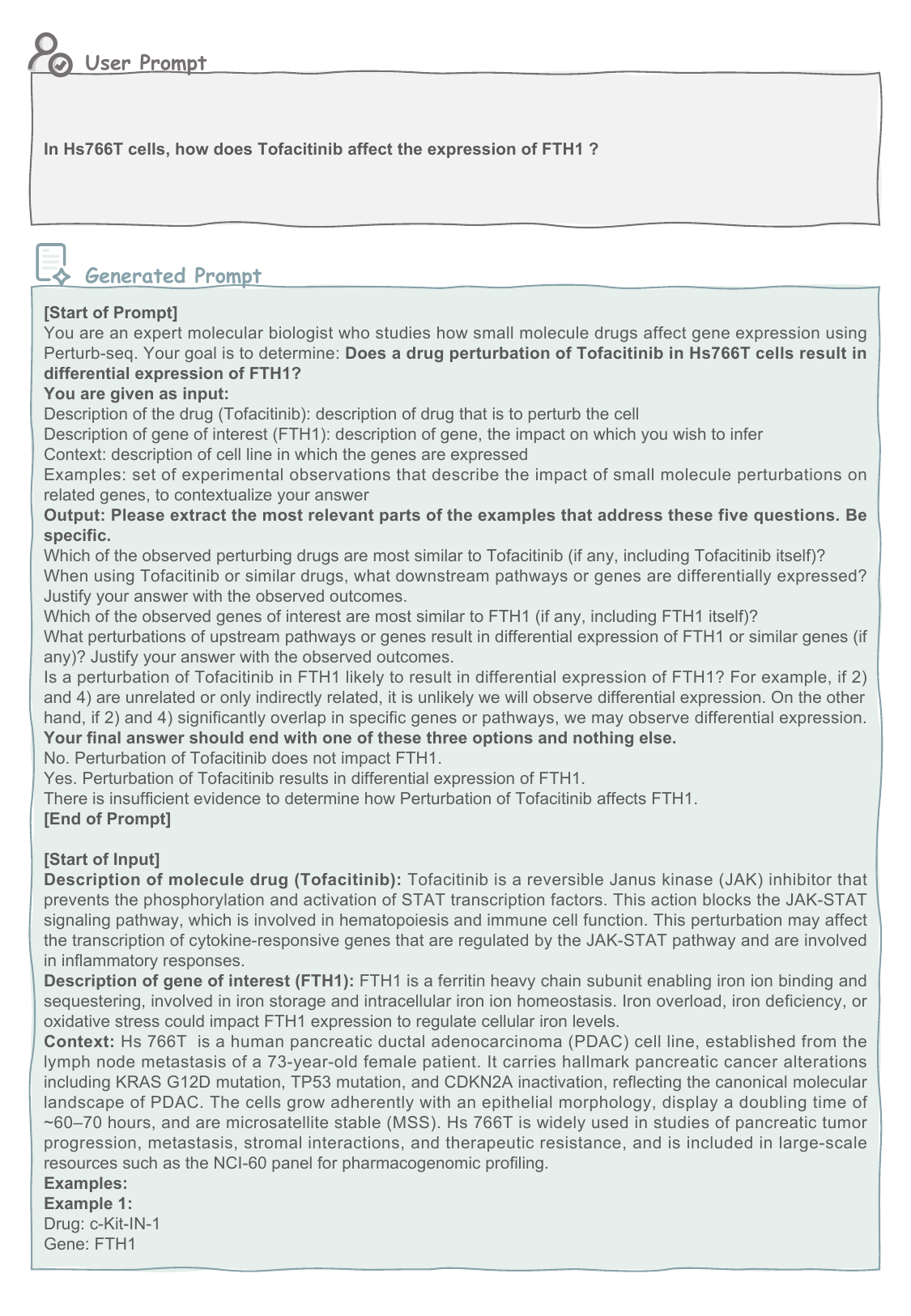}

    \label{}
\end{figure}
\begin{figure}
    \centering
    \includegraphics[width=1\linewidth]{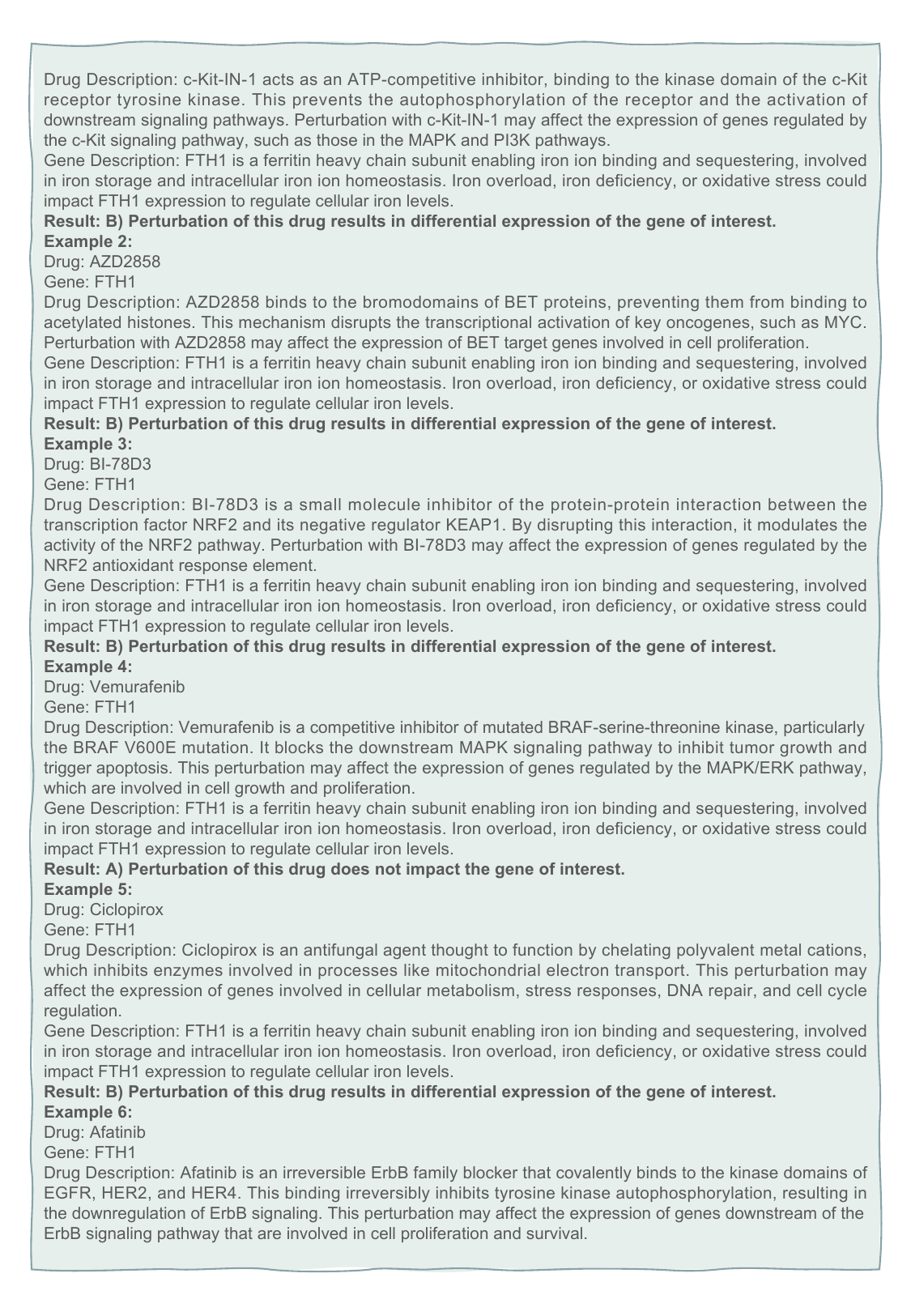}

    \label{}
\end{figure}
\begin{figure}
    \centering
    \includegraphics[width=1\linewidth]{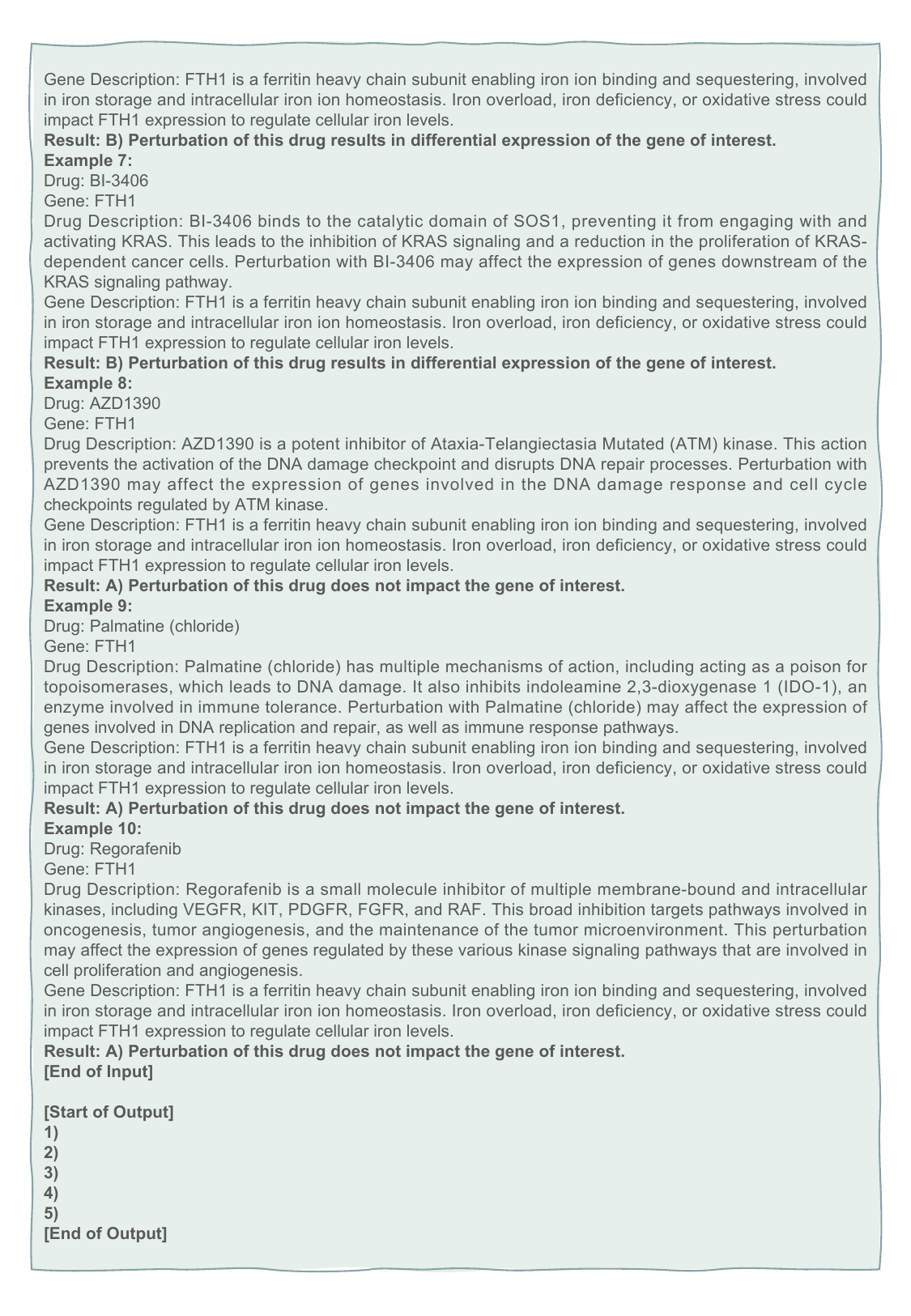}

    \label{}
\end{figure}
\begin{figure}
    \centering
    \includegraphics[width=1\linewidth]{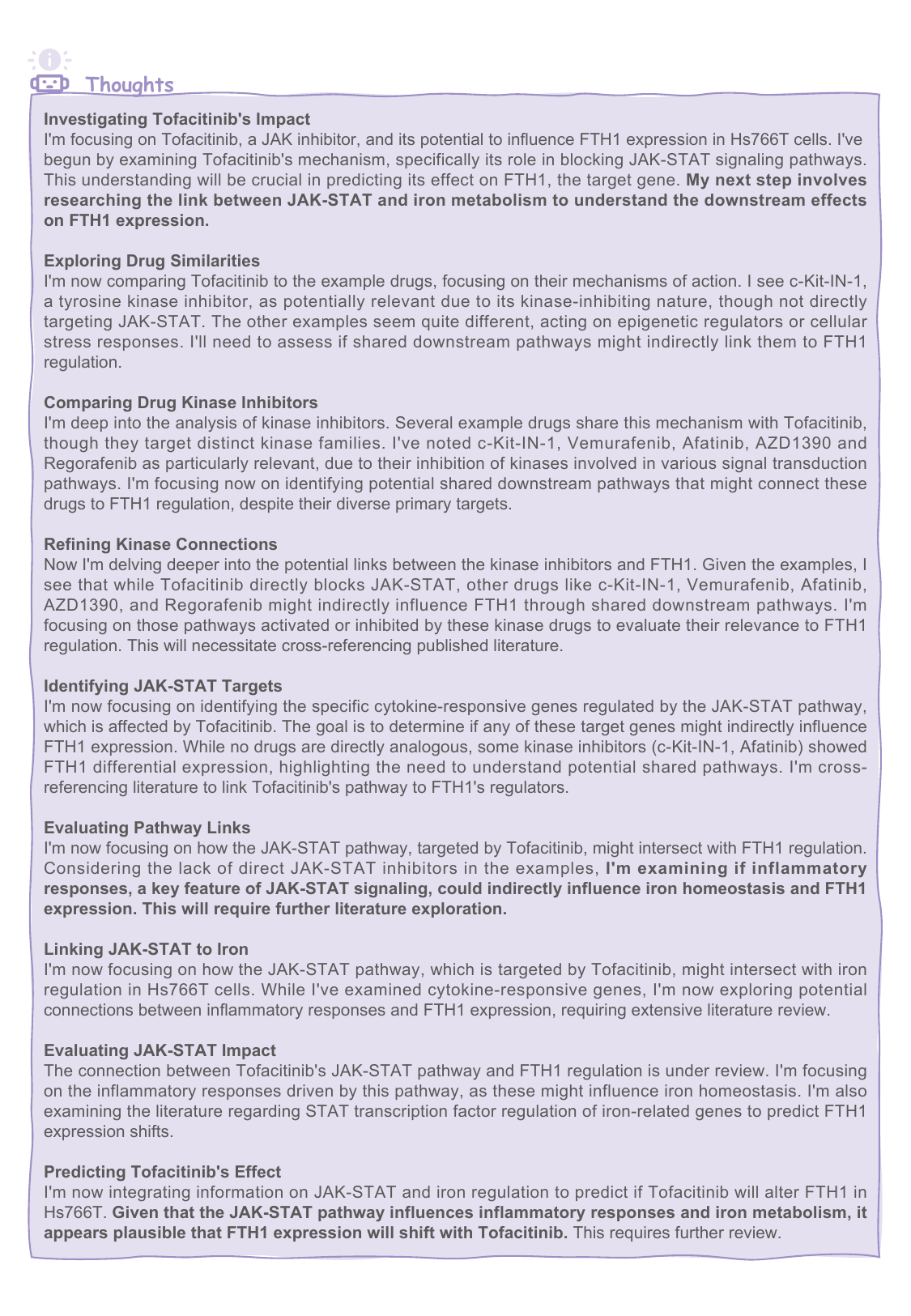}

    \label{}
\end{figure}
\begin{figure}
    \centering
    \includegraphics[width=1\linewidth]{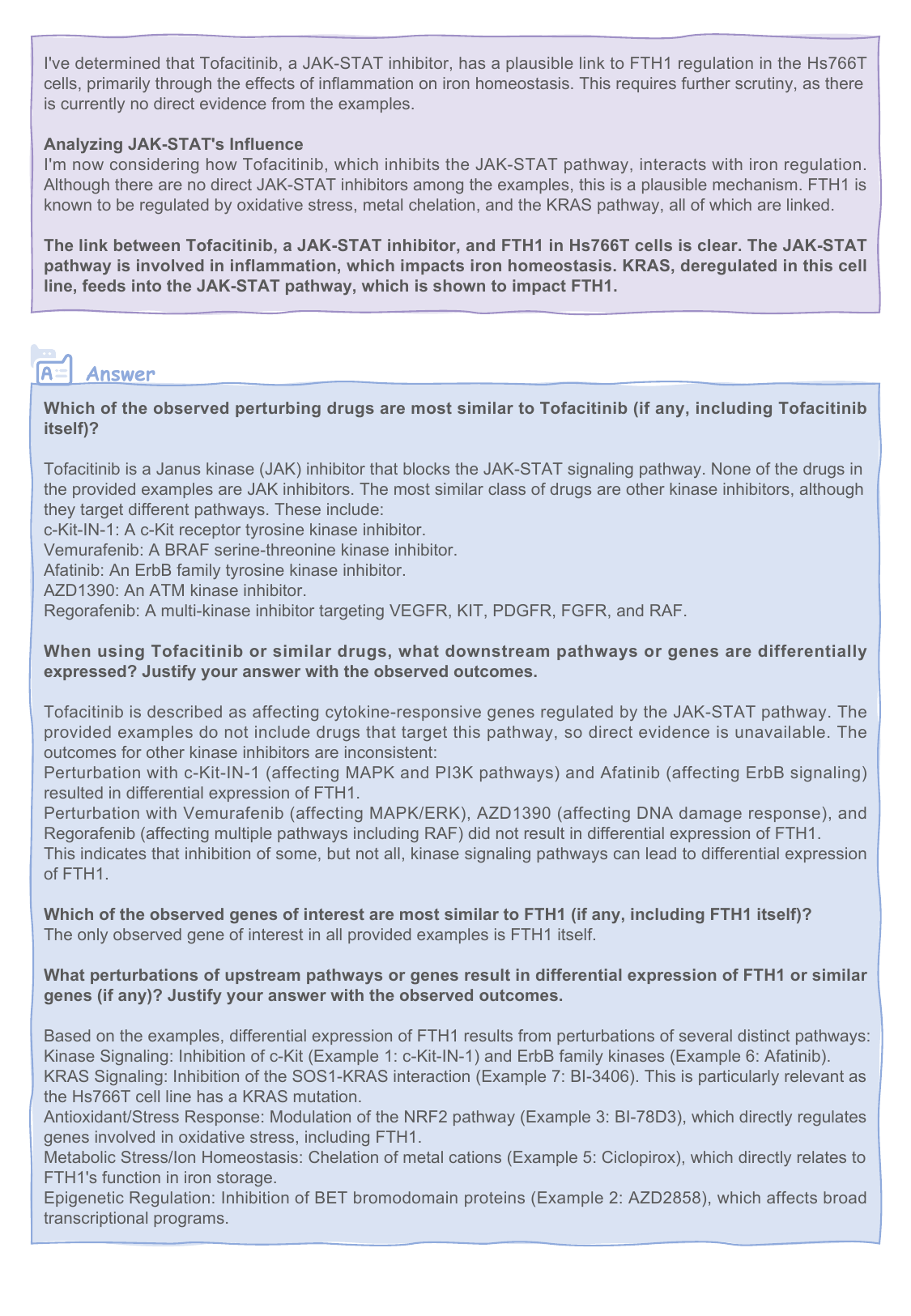}
    
   \
    \label{}
\end{figure}
\begin{figure}
    \centering
    \includegraphics[width=1\linewidth]{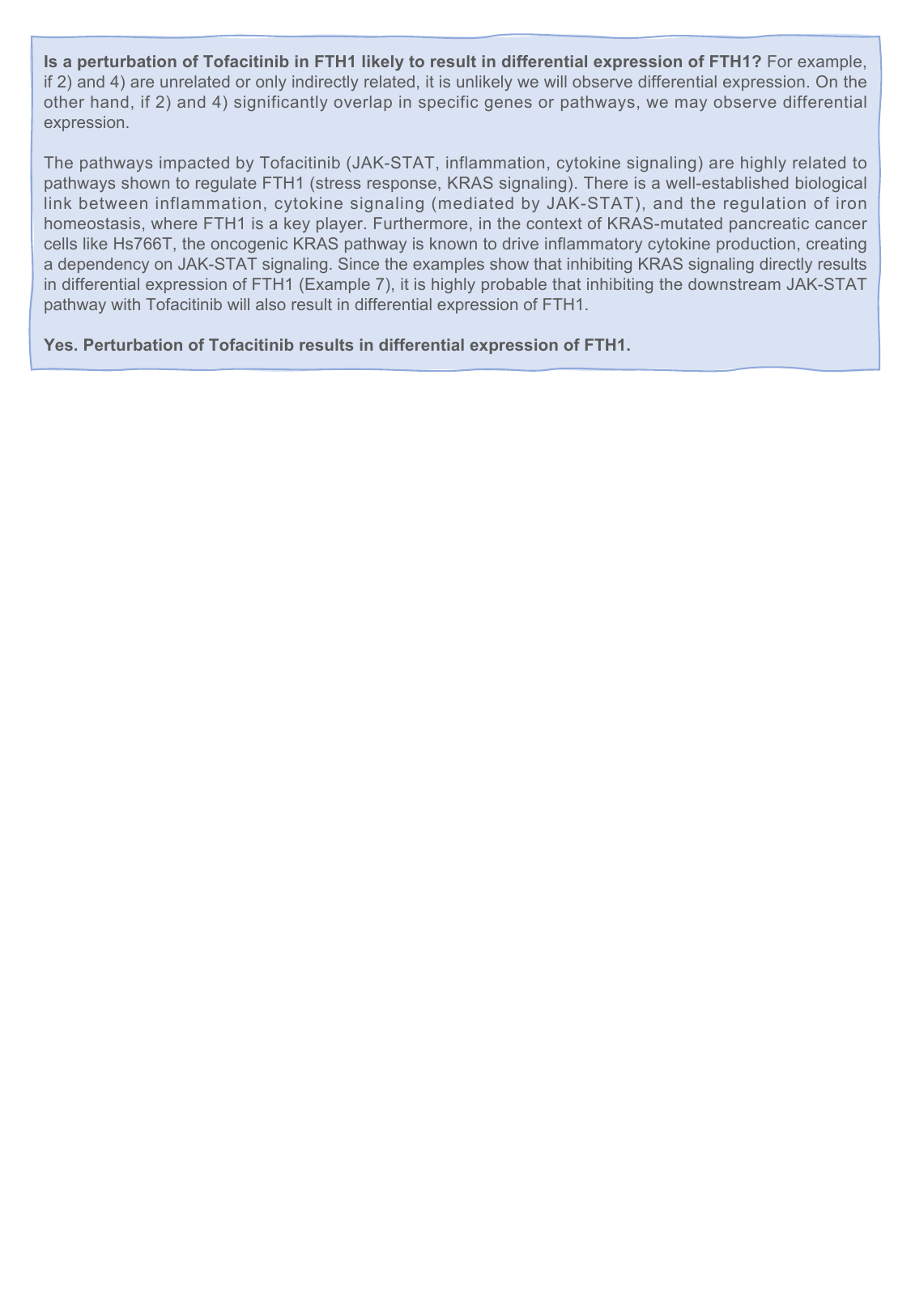}

    \label{}
\end{figure}
\label{appP}
\begin{figure}
    \centering
    \includegraphics[width=1\linewidth]{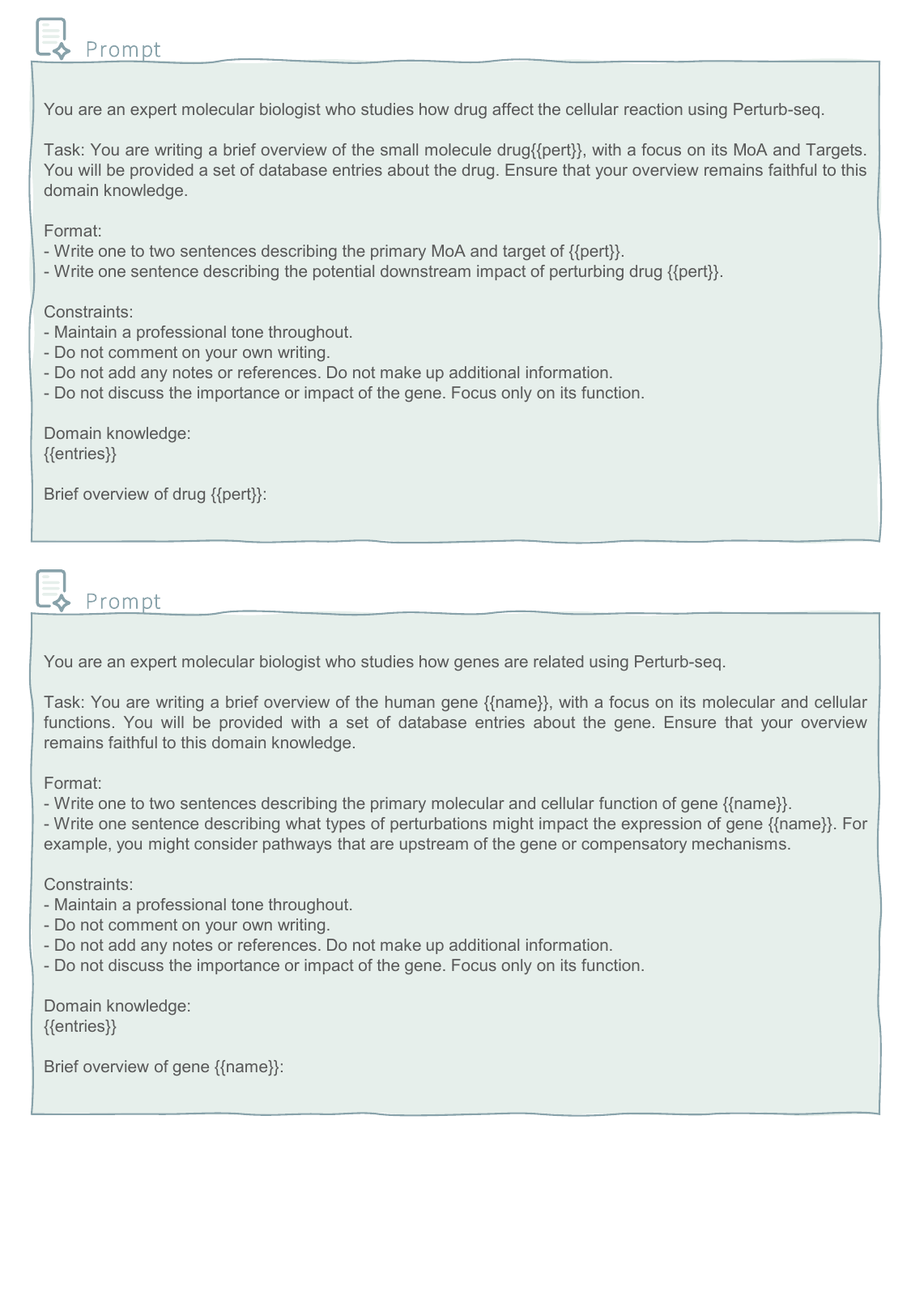}

    \label{}
\end{figure}
\begin{figure}
    \centering
    \includegraphics[width=1\linewidth]{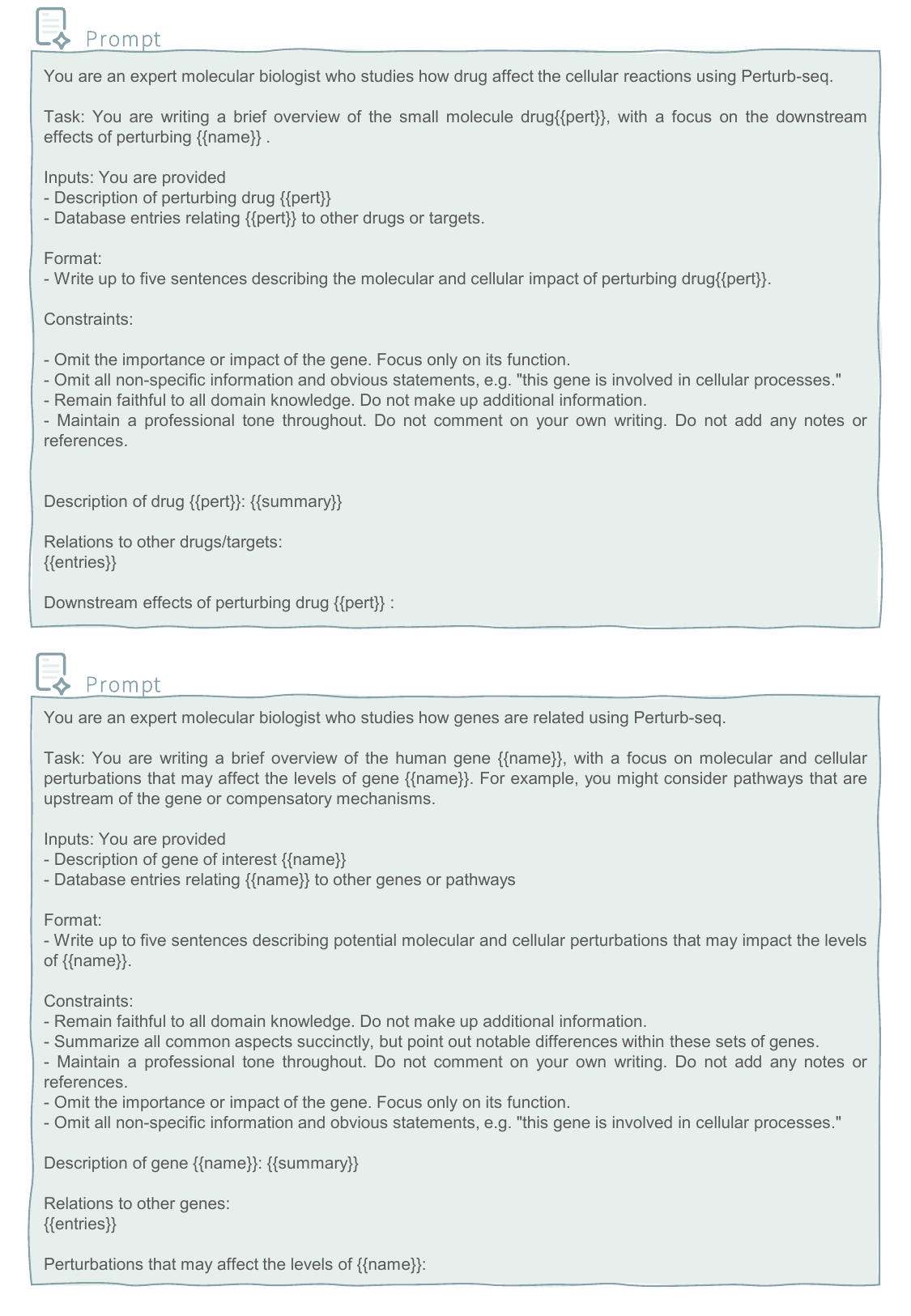}

    \label{}
\end{figure}

\end{document}